\documentclass[11pt]{article} 
\pdfoutput=1 

\usepackage{jheppub} 

\usepackage[utf8]{inputenc}
\usepackage{ulem}
\usepackage[english]{babel}
\usepackage{amsmath,amssymb,amsbsy,amstext,amsthm,booktabs,simplewick,exscale,relsize,slashed,graphicx,amsfonts,upgreek, xcolor,subcaption}
\usepackage{multirow,color,dcolumn,bm,enumerate}
\usepackage{hyperref}
\usepackage{dsfont}
\usepackage{ragged2e} 
\DeclareUnicodeCharacter{2212}{-}
\usepackage{threeparttable}
\usepackage{makecell}

\usepackage{hyperref}

\newcommand{\sgw}{\sigma_{w}}
\newcommand{\lw}{L_{w}}

\newcommand{\bsk}{\boldsymbol{k}}
\newcommand{\bsq}{\boldsymbol{q}}
\newcommand{\bsm}{\boldsymbol{m}}
\newcommand{\ma}{\mathcal{A}}
\newcommand{\mcp}{\mathcal{P}}
\newcommand{\md}{\mathrm{d}}

\title{Scalar-induced gravitational wave from domain wall perturbation}

\author[a,b,1]{Bo-Qiang Lu}
\emailAdd{bqlu@zjhu.edu.cn}

\affiliation[a]{School of Science, Huzhou University, Huzhou, Zhejiang 313000, China}
\affiliation[b]{Zhejiang Key Laboratory for Industrial Solid Waste Thermal Hydrolysis Technology and Intelligent Equipment}

\abstract{
Domain walls represent two-dimensional topological defects that emerge from the spontaneous breaking of discrete symmetries in various new physics models. 
In this study, we undertake the first calculation of gravitational waves produced by scalar perturbations generated from the gravitational wave network. 
Our findings indicate that the gravitational wave spectrum is notably distinct from that of other sources. 
This opens up a promising avenue for future gravitational wave experiments aimed at exploring the role of domain walls in the early universe. 
}

\begin{document}
\maketitle

\setcounter{page}{2}

\section{Introduction}\label{sec:intro}
The detection of gravitational waves (GWs) from binary black hole mergers by LIGO in 2015 marked the dawn of GW astronomy~\cite{LIGOScientific:2016aoc}.
Recent pulsar timing array (PTA) observations have further provided compelling evidence for a stochastic GW background in the nanohertz frequency range~\cite{NANOGrav:2023gor, EPTA:2023fyk, Reardon:2023gzh, Xu:2023wog}.
Owing to the weak coupling between GWs and matter, GWs produced in the early universe propagate virtually unimpeded, free from absorption or scattering by the primordial plasma.
This unique property renders GWs a powerful probe of cosmic history prior to Big Bang Nucleosynthesis (BBN), a period largely inaccessible to electromagnetic observations.
Potential sources of such primordial GWs include first-order phase transitions, quantum fluctuations amplified during inflation, oscillations of cosmic-string networks, and the collapse of domain walls (DWs), among others (see Ref.~\cite{Caprini:2018mtu} for a comprehensive review).

Scalar-induced GWs represent a generic production mechanism whereby second-order tensor perturbations are sourced by primordial curvature perturbations, even though scalar and tensor modes remain linearly decoupled at first order in cosmological perturbation theory (see Ref.~\cite{Domenech:2021ztg} for a recent review).  
A widely studied source of scalar-induced GWs involves the re-entry of superhorizon scalar perturbation modes generated during inflation. At the cosmic microwave background (CMB) scale, observations constrain the nearly scale-invariant primordial scalar power spectrum to $\sim 10^{-9}$~\cite{Planck:2018nkj,Byrnes:2018txb}, suggesting a second-order GW power spectrum of $\sim 10^{-18}$—far below current experimental sensitivity~\cite{Ananda:2006af}. However, scale-dependent scalar spectra, such as those arising in multi-field inflation~\cite{Ananda:2006af,Bugaev:2009zh,Alabidi:2012ex}, or scenarios where density perturbations grow during dust-dominated eras (e.g., via primordial black hole (PBH) formation~\cite{Assadullahi:2009nf,Alabidi:2013lya,Kovetz:2017rvv,Kohri:2018awv,Inomata:2019ivs,Domenech:2020ssp}), can amplify second-order GWs. In PBH scenarios, initially isocurvature perturbations transition to curvature perturbations in the dust-dominated phase~\cite{Domenech:2020ssp}, potentially enhancing GW signals to levels comparable with first-order predictions~\cite{Kohri:2018awv}.  
In this work, we begin by computing the curvature perturbations produced by the DW network. Building on these results, we then derive the power spectrum of second-order GWs induced by these curvature perturbations. 
Crucially, as the DW network's energy fraction increases over time, the resulting scalar-induced GW signal can become observationally significant, provided the DWs persist long enough after their formation.

DWs are two-dimensional topological defects formed during the spontaneous breaking of a discrete symmetry. In the scaling regime, their energy density evolves as \(\rho_w = \sigma_{w} / t\), where \(\sigma_{w}\) denotes the wall’s surface tension. By contrast, the radiation energy density scales as \(\rho_{\text{rad}} = 3/(32\pi G t^2)\). When \(t \sim 1/(G\sigma_{w})\), DW energy density begins to dominate the universe, resulting in the longstanding DW problem. 
A standard resolution involves introducing a discrete symmetry-breaking operator to generate a bias potential. Annihilation of DWs commences once this potential’s energy density becomes comparable to \(\rho_w\).  
Beyond bias potentials, several mechanisms can mitigate or resolve this issue: 
\begin{itemize}
    \item[1.] Quantum tunneling decay: Metastable DWs may decay via quantum tunneling, as proposed in early studies~\cite{Kibble:1982dd,Preskill:1992ck}. 
    \item[2.] Network disruption: Nucleation of string-bounded holes or collapse of string-bounded DW networks can fragment the structure~\cite{Dunsky:2021tih}.
    \item[3.] Temperature-dependent tension: Walls with tension decreasing at higher temperatures may melt into the plasma~\cite{Babichev:2021uvl,Babichev:2023pbf}.  
    \item[4.] Embedded walls: Configurations stabilized by plasma effects despite lacking topological stability in vacuum~\cite{Schroder:2024gsi}.
\end{itemize}


If DWs persist long enough to constitute \(\sim 10\%\) of the total energy density, their dynamical evolution can imprint observable cosmological signatures~\cite{Lu:2023mcz,Gouttenoire:2023ftk,Blasi:2023sej}. Notably, DWs generated at the \(\sim 200\ \text{TeV}\) scale—annihilated via QCD-instanton-induced bias potentials during the QCD phase transition—provide a natural explanation for recent PTA signals, including the NANOGrav 15-year dataset~\cite{NANOGrav:2023gor} and IPTA-DR2~\cite{Antoniadis:2022pcn,Chiang:2020aui,Lu:2023mcz}.  
Recent work by Chiang, Li, and the author demonstrates that Poisson fluctuations in DW networks seed horizon-sized overdensities, enabling PBH formation when DWs contribute \(\sim 10\%\) of the energy density~\cite{Lu:2024ngi,Lu:2024szr}. This mechanism’s efficiency suggests that scalar perturbations sourced by DWs could similarly generate detectable GWs~\cite{Saito:2008jc,Saito:2009jt}. 
In this work, we present a detailed calculation of the scalar-induced GW spectrum from DW perturbations. A companion study will assess detection prospects for these GW signals in next-generation space-based interferometers.  

This work is organized as follows. In section~\ref{sec:perturb}, we examine the curvature perturbation generated by the DW. Section~\ref{sec:IGW} presents the general formulations for GW induced by scalar perturbations at second order. Subsequently, in section~\ref{sec:GWfromDW}, we derive the spectrum of induced GWs resulting from the DW perturbation. Finally, section~\ref{sec:conclusions} provides a summary of our findings and conclusions. 
Additionally, in appendix~\ref{app:splength}, we elucidate the physical significance of the separation length of the DW. In appendix~\ref{app:DWevolution}, we present the formulas relevant to the evolution of the correlation length of the DW.

\section{Perturbations from DW}\label{sec:perturb}
In this section, we will first calculate the first-order scalar perturbations in the metric induced by the DW. Subsequently, we will analyze the power spectrum of the curvature perturbations associated with the generated GWs.

\subsection{The curvature perturbation by DW}
Consider the linearly perturbed Friedmann-Lemaître-Robertson-Walker (FLRW) metric, which incorporates a second-order tensor perturbation \( h_{ij} \) in the Newtonian gauge within a spatially flat universe.
\begin{equation}
    \mathrm{d} s^2=a^2(\eta)\left\{-(1+2 \Psi) \mathrm{d} \eta^2+\left[(1+2 \Phi) \delta_{i j}+\frac{h_{i j}}{2}\right] 
    \mathrm{d} x^i \mathrm{~d} x^j\right\} ,
\end{equation}
where the conformal time \( \eta \) is related to the cosmic time by the equation \( dt = a d\eta \). The Newtonian potential \( \Psi \) and the intrinsic curvature perturbation \( \Phi \) are gauge-invariant quantities, formed from combinations of metric perturbations, and \( h_{ij} \) represents the second-order tensor mode. We disregard the first-order tensor perturbation and vector perturbations. 
The tensor mode adheres to the tracelessness condition \( h_i^i = 0 \) and the transversality condition \( \partial^i h_{ij} = 0 \). The tracelessness condition ensures that tensor perturbations do not affect the trace of the metric perturbation, while the transversality condition guarantees that the tensor perturbations are divergence-free and propagate as transverse waves.

During the radiation era, the scale factor, Hubble rate, and energy density evolve according to the relationships 
\begin{equation}
    a(\eta) = a_i \frac{\eta}{\eta_i},~~\mathcal{H} = aH = \eta^{-1},~~{\rm and~~} \rho_{\rm rad} \propto \eta^{-4}.
\end{equation}
It is important to note that second-order perturbation theory encounters the so-called gauge issue: the theory is not gauge invariant under gauge transformations. Consequently, the derived tensor spectrum is dependent on the choice of gauge (see Ref.~\cite{Domenech:2021ztg} for a comprehensive review). 
Interestingly, previous investigations have revealed that in a radiation-dominated universe, the numerical results for the induced tensor spectrum in both the Newtonian gauge and the synchronous gauge are found to be identical~\cite{DeLuca:2019ufz, Inomata:2019yww, Yuan:2019fwv}.

Consider a single-component universe dominated by static DWs. The gauge-invariant scalar perturbations \(\tilde{\Phi}\) and \(\tilde{\Psi}\) obey the linearized Einstein equations~\cite{Vilenkin:1984ib,Nambu:1990ez}:
\begin{eqnarray}
    \label{eq:Einstein1}
    \nabla^2 \tilde{\Phi} &=& -4 \pi G a^2\bar{\rho} \delta, \\
    \label{eq:Einstein2}
    \nabla^2(\tilde{\Phi} + \tilde{\Psi}) &=& 8 \pi G a^2 \bar{p} \Pi,
\end{eqnarray}
where $G$ is the Newton's gravitational constant, \(\bar{\rho}\) and \(\bar{p}\) are the background energy density and pressure, \(\delta\) is the relative density perturbation, and \(\Pi\) is the anisotropic stress. Tildes denote perturbations in a universe exclusively containing DWs.
For an idealized thin DW (neglecting finite thickness) aligned with the \(yz\)-plane at comoving position \(x=0\), the energy-momentum tensor in the wall frame is~\cite{Vilenkin:1984ib}:
\begin{equation}
    T_{\mu\nu} = \sigma_{w} \delta(ax) \, \text{diag}(1, 0, -1, -1),
\end{equation}
where \(\sigma_{w}\) is the wall’s surface tension, and \(\delta(ax)\) is the dimensionless Dirac delta function. This gives \(\bar{p}\Pi = -\sigma_{w}\delta(ax)\), reducing Eqs.~\eqref{eq:Einstein1} and~\eqref{eq:Einstein2} to:
\begin{eqnarray}
    \label{eq:Poisson1}
    \nabla^2\tilde{\Phi}(x) &=& -4\pi Ga^2\sigma_{w}\delta(ax), \\
    \label{eq:Poisson2}
    \nabla^2\left[\tilde{\Psi}(x) + \tilde{\Phi}(x)\right] &=& -8\pi Ga^2\sigma_{w}\delta(ax).
\end{eqnarray}
From Eqs.~\eqref{eq:Poisson1} and~\eqref{eq:Poisson2}, we derive two key results:  
\begin{itemize}
    \item[1.] Repulsive Gravity: Unlike non-relativistic matter, which generates attractive gravitational effects, DWs produce repulsive forces~\cite{Vilenkin:1984ib}.
    \item[2.] Potential Equality: Combining these equations yields \(\tilde{\Psi}(x) = \tilde{\Phi}(x)\) for DW-dominated perturbations. 
\end{itemize}

To solve for the scalar potential \(\tilde{\Psi}\), we note from Eq.~\eqref{eq:Poisson1} that \(\tilde{\Psi}\) must be continuous at \(x=0\) while its derivative exhibits a jump discontinuity. This condition implies \(\tilde{\Psi}(x)\) depends on \(|x|\). Imposing parity symmetry (\(\tilde{\Psi}(x) = \tilde{\Psi}(-x)\)) and asymptotic behavior (\(\tilde{\Psi}(x) \propto 1/|x|\) at large \(|x|\)), the solution becomes~\cite{Nambu:1990ez}:
\begin{equation}\label{eq:pot1}
    \tilde{\Psi}(x) = 2\pi G a(\eta)\sigma_{w} \left(\sqrt{x^2 + d_w^2} - |x|\right),
\end{equation}
where \(d_w\) represents the wall’s comoving characteristic length. In a wall network, \(d_w\) corresponds to the average separation between walls, related to the physical correlation length \(L_w\) by \(d_w = L_w/a(\eta)\). Detailed derivations of this relationship and its connection to wall dynamics are provided in appendix~\ref{app:splength} (see also Ref.~\cite{Lu:2024ngi}).
For a periodic array of walls at positions \(x_j = jd_w\) (\(j = 0, \pm1, \pm2, \ldots\)), the total potential can be approximated as:
\begin{equation}\label{eq:pot2}
    \tilde{\Psi}(x) = 2\pi G a(\eta)\sigma_{w} \sum_j \left(\sqrt{(x-jd_w)^2 + d_w^2} - |x-jd_w|\right) \simeq C\cos^2\left(\frac{\pi x}{d_w}\right),
\end{equation}
where the coefficient \(C\) is determined by matching the spatial averages of both potential forms, yielding \(C \simeq \pi G\sigma_{w}L_w\). 
The Fourier components of this potential are:
\begin{equation}\label{eq:potk1}
    \tilde{\Psi}_k(\eta) = \frac{C}{2}\sqrt{\frac{\pi}{2}} \, \delta\left(k - \frac{2\pi}{d_w}\right),
\end{equation}
where \(k\) denotes the comoving wavenumber. We adopt the convention \(a(\eta_0) = 1\) for the current scale factor of the universe, with \(\eta_0 = 1.47 \times 10^{18}\)~s representing today’s conformal time.


In this study, we will examine a two-component universe that is predominantly dominated by radiation, wherein DWs are generated through the spontaneous breaking of a discrete symmetry. It is necessary for these DWs to be annihilated by a biased potential prior to their dominance in the universe. 
Within this framework, the variables \( \Phi \) and \( \Psi \) will be utilized to denote the values of the perturbations in the two-component universe. The anisotropic stress for this universe can be expressed as 
\begin{equation}
    \Pi = f_{\gamma} \Pi_{\gamma} + f_w \Pi_w
\end{equation}
where \( f_{\gamma} = \bar{\rho}_{\rm rad}/\bar{\rho}_{\rm tot} \simeq 1 \) and $f_w=\bar{\rho}_w/\bar{\rho}_{\rm tot}\simeq \bar{\rho}_w/\bar{\rho}_{\rm rad}$ represent the energy density fractions associated with radiation and the DWs, respectively. 
Here, \( \Pi_{\gamma} \) and \( \Pi_w \) denote anisotropic stresses from radiation and DWs, respectively. Given that radiation functions as a perfect fluid, its anisotropic stress is identically zero, specifically \( \Pi_{\gamma} = 0 \). Therefore, the anisotropic stress expressed in Equation \eqref{eq:Einstein2} for the two-component universe can be simplified to \( \Pi = f_w(\eta) \Pi_w \).

In the limit where \( f_w(\eta) \to 0 \) (which corresponds to scenarios in which the DWs are absent or have decayed), the anisotropic stress is eliminated, leading to the condition \(\Phi(x) = -\Psi(x)\) as indicated by the linearized Einstein equations. Additionally, from the Einstein equations, we infer that \(\tilde{\Phi}(x) = \tilde{\Psi}(x)\) in a DW-dominated universe, which corresponds to the case where \( f_w(\eta) = 1 \). 
Moreover, the perturbation introduced by DWs is isocurvature at the initial time \( \eta_i \), meaning that the density perturbation satisfies \(\delta \rho(\eta_i) = 0\). Consequently, we have \(\Phi(\eta_i) = 0\) since the curvature perturbation is directly proportional to the density perturbation.
Thus, we can estimate the curvature perturbation due to DWs in a two-component universe dominated by radiation as follows:
\begin{equation}
    \Phi(\eta,x)\simeq f_w(\eta)\tilde{\Phi}(\eta,x)=f_w(\eta)\tilde{\Psi}(\eta,x).
\end{equation}
It is important to note that this relation should not be applied to scenarios occurring after the annihilation of the DWs.

\subsection{The power spectrum of the curvature perturbation}

It is noteworthy that the potential, along with the power spectrum derived from Eq.~\eqref{eq:potk1}, is characterized as one-dimensional (with the power spectrum defined in Eq.~\eqref{eq:Phipw}). In the context of isotropic perturbations, one can transition from the one-dimensional power spectrum \( \Delta_{\rm 1D}(k) \) to the three-dimensional power spectrum \( \Delta_{\rm 3D}(k) \) using the relationship outlined in~\cite{Kaiser:1990xe}:
\begin{equation}\label{eq:D3D}
    \Delta_{\rm 3D}(k)=\frac{1}{k^2}\int_{k}^{\infty}\Delta_{\rm 1D}(y)ydy.
\end{equation}
By employing the one-dimensional potential as indicated in Eq.~\eqref{eq:potk1} and utilizing Eq.~\eqref{eq:D3D}, while focusing on perturbations that are horizon-sized, we can derive the three-dimensional isotropic power spectrum for the wall:
\begin{equation}
    \mathcal{P}_{\tilde{\Psi}}(k, \eta)=\frac{C^2k^2}{16\pi}\delta\left(k-\frac{2\pi}{d_w}\right).
\end{equation}
Therefore, we can observe that the three-dimensional potential in momentum space can be formulated from the one-dimensional potential in Eq.~\eqref{eq:potk1} by making the substitution \(\tilde{\Psi}_k(\eta) \to \tilde{\Psi}_k(\eta) / \sqrt{k}\).
In light of these findings, the Fourier component of the curvature perturbation within the two-component universe can be articulated as follows:
\begin{eqnarray}\label{eq:cp}
    \Phi_k(\eta)=\frac{Cf_{w}(\eta)}{2}\sqrt{\frac{\pi}{2k}}\delta\left(k-\frac{2\pi}{d_w}\right).
\end{eqnarray}

Equation \eqref{eq:cp} demonstrates that the DW network serves as a continuous source of curvature perturbation, with the evolution of the power spectrum being encapsulated in the correlation length \( L_w = a(\eta) d_w \) of the wall. Detailed information regarding the physical meaning of $L_w$ and the evolution of \( L_w \) with respect to cosmic time can be found in appendix~\ref{app:splength} and appendix~\ref{app:DWevolution}. 
After a period of rapid expansion, the DW transitions into a scaling regime, in which \( L_w \) scales with cosmic time as \( L_w \simeq t/\ma \), where \( \ma \simeq 1 \) is derived from simulations~\cite{Hiramatsu:2013qaa}. The energy density of the wall is connected to the correlation length through the one-scale assumption, expressed as \( \rho_w = \sigma_w/L_w \) (for further details on the velocity-dependent one-scale model, please refer to  appendix~\ref{app:DWevolution} and Refs.~\cite{Martins:2016book, Martins:2016ois}). 
In a radiation-dominated universe, the scale factor can be represented as \( a(\eta) = a_i \eta/\eta_i \), where \( \eta_i \) denotes the reference initial time. The cosmic time is given by \( dt = a(\eta) d\eta \), and we can approximate \( t \simeq a_i \eta^2/(2\eta_i) \). Consequently, the correlation length is expressed as:
\begin{equation}
    \lw=\frac{a_i\eta^2}{2\ma\eta_i}~~{\rm and}~~d_w=\frac{\lw}{a(\eta)}=\frac{\eta}{2\ma}.
\end{equation}
The energy fraction of the wall is then defined by:
\begin{equation}
    f_w(\eta)=\frac{\bar{\rho}_w}{\bar{\rho}_{\rm rad}}=\frac{32}{3}\pi \ma G\sgw t=\frac{32\pi \ma G\sgw}{3}\frac{a_i\eta^2}{2\eta_i}.
\end{equation}
Following common practice in the literature, we decompose the potential as follows:
\begin{equation}\label{eq:Phidp}
    \Phi_{k}(\eta)=T(x)\phi_{k},
\end{equation}
where \( \phi_{k} \) represents the value of the potential at the initial time \( \eta_i \), and \( T(x) \) is a transfer function that captures the changes in the amplitudes of the potential between \( \eta_i \) and \( \eta \). For our analysis, we adopt the notation \( x \equiv k\eta \).
For our specific case, the two functions are given by:
\begin{equation}\label{eq:Tphi}
    T(x)=x^5\delta\left(x-4\pi\ma\right)~~{\rm and}~~\phi_k=\frac{B}{k^5\sqrt{k}},
\end{equation}   
where
\begin{equation}
    B=\frac{2^{\frac{3}{2}}\pi^{\frac{5}{2}}a_i^2G^2\sgw^2}{3\eta_i^2}.
\end{equation}

Assuming Gaussian fluctuations, the Bardeen potential can be expressed as \(\Phi_{\bsk} (\eta) = \Phi_k(\eta) \hat{E}(\bsk)\)~\cite{Ananda:2006af}, where \(\hat{E}\) is a Gaussian random variable with unit variance, possessing the property 
\begin{equation}
    \left\langle \hat{E}^*\left(\bsk_1\right) \hat{E}\left(\bsk_2\right) \right\rangle = \delta^3\left(\bsk_1 - \bsk_2\right).
\end{equation}
The reduced power spectrum $\mathcal{P}_{\Phi}(k, \eta)$ for scalar perturbations is then defined by the relation
\begin{equation}\label{eq:Phipw}
    \left\langle\Phi^*\left(\eta\right) \Phi\left(\eta\right)\right\rangle=\frac{2 \pi^2}{k^3} 
    \delta\left(\bsk_1-\bsk_2\right) \mathcal{P}_{\Phi}(k, \eta).
\end{equation}

Utilizing the decomposition presented in Eq.~\eqref{eq:Phidp}, the potential is given by \(\Phi_{\bsk}(\eta) = T(k\eta) \phi_{\bsk}(k)\), where \(\phi_{\bsk} = \phi_k \hat{E}(\bsk)\). Consequently, the power spectrum can be expressed as:
\begin{equation}\label{eq:mcp}
    \mcp_{\phi}(k)=\frac{B^2}{2\pi^2k^8}= \left( \frac{k_B}{k} \right)^8,
\end{equation}
with 
\begin{equation}
    k_B \simeq \left(\frac{G\sigma_w}{\eta_0}\right)^{1/2}.
\end{equation}
This indicates that the power spectrum scales as \(k^{-8}\), emphasizing its significance at larger scales. It is anticipated that the scalar-induced GWs resulting from DW perturbations can experience substantial enhancement in the later stages of cosmic evolution.
By employing the time-temperature relation:
\begin{equation}
    t = \left( \frac{90}{32\pi^3 g_*(T)} \right)^{1/2} \frac{1}{G^{1/2} T^2},
\end{equation}
where \(g_*(T)\) represents the effective number of relativistic degrees of freedom in the plasma, and the conformal Hubble parameter is defined as \(\mathcal{H} = 1/\eta \simeq 1/(2t \eta_0)^{1/2}\), we derive:
\begin{equation}
    \left( \frac{k_B}{\mathcal{H}} \right)^2 = \frac{2\sigma_w}{m_{\rm pl} T^2} \left( \frac{90}{32\pi^3 g_*(T)} \right)^{1/2},
\end{equation}
where \(m_{\rm pl} = G^{-1/2} = 1.22 \times 10^{19}~\mathrm{GeV}\) is the Planck mass.  

Crucially, the energy density of DWs becomes cosmologically dominant when \(\rho_w(T_{\rm dom}) \simeq \rho_{\rm tot}(T_{\rm dom}) = 3H^2/(8\pi G)\). This condition yields the dominant temperature:
\begin{equation}\label{eq:Tdom}
    T_{\rm dom} = \left( \frac{8\sqrt{5} \mathcal{A} \sigma_w}{\pi^{1/2} g_*^{1/2}(T) m_{\rm pl}} \right)^{1/2} \simeq 282.8~\mathrm{GeV} \times \left( \frac{\mathcal{A}}{1.0} \right)^{1/2} \left( \frac{107}{g_{*}(T)} \right)^{1/4} \left( \frac{\sigma_w^{1/3}}{10^8~\mathrm{GeV}} \right)^{3/2}.
\end{equation}

Causality imposes that the power spectrum of curvature perturbations arises only when DWs are confined within the cosmological horizon, satisfying \(k = 2\pi/d_w \sim \mathcal{H}\). This condition introduces a lower cutoff on the power spectrum generated by DW perturbations:  
\begin{equation}  
    \mathcal{P}_{\phi,\mathrm{cut}} \sim \left( \frac{k_B}{\mathcal{H}} \right)^8 \lesssim \left( \frac{3}{16\pi \ma} \right)^8 \sim 10^{-10}.  
\end{equation}  
Here, the inequality reflects the requirement that DWs annihilate prior to their cosmological domination (\(T > T_{\rm dom}\)).  
Consequently, causality prevents infrared (IR) divergence in the power spectrum, provided the DWs annihilate before dominating the universe’s energy density.

\section{Induced GW}\label{sec:IGW}
The spatial correlations of the curvature perturbation act as the source for the induced GWs. In this section, we will follow the methodologies outlined in Refs.~\cite{Ananda:2006af, Baumann:2007zm, Kohri:2018awv, Papanikolaou:2020qtd} to calculate the GWs that are induced by scalar perturbations at second order.

\subsection{The tensor mode of the induced GW}
\begin{figure}[t!]
    \centering 
    \includegraphics[width=0.40\textwidth]{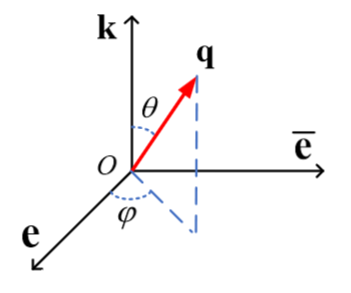}
    \caption{Momentum transfer vector \(\boldsymbol{q}\) in the \(\{e, \bar{e}, \boldsymbol{k}\}\) coordinate system.}
    \label{fig:sc}
\end{figure}

The evolution of the tensor mode is governed by~\cite{Ananda:2006af}
\begin{equation}\label{eq:hxEOM}
    h_{ij}^{\prime \prime}+2 \mathcal{H} h_{ij}^{\prime}-\nabla^2 h_{ij}=-4 \hat{\mathcal{T}}_{ij}{ }^{l m} \mathcal{S}_{lm},
\end{equation}
where the prime denotes a derivative with respect to conformal time. The source term is given by~\cite{Ananda:2006af}
\begin{equation}
    \mathcal{S}_{ij}=4 \Phi\Phi_{\mid ij}+2 \Phi_{\mid i} \Phi_{\mid j}-\frac{3}{8\pi G a^2 \rho}\left[\mathcal{H}^2 
    \Phi_{\mid i} \Phi_{\mid j}+2 \mathcal{H} \Phi_{\mid i} \Phi_{\mid j}^{\prime}+\Phi_{\mid i}^{\prime} 
    \Phi_{\mid j}^{\prime}\right],
\end{equation}
where the pipe denotes the spatial covariant derivative~\cite{Ananda:2006af}. 
Transforming Eq.~\eqref{eq:hxEOM} to Fourier space yields:  
\begin{equation}\label{eq:hkEOM}  
    h_{\boldsymbol{k}}^{s, \prime \prime} + 2 \mathcal{H} h_{\boldsymbol{k}}^{s, \prime} + k^2 h_{\boldsymbol{k}}^s = 4 S_{\boldsymbol{k}}^s,  
\end{equation}  
where \( s = +, \times \) labels the GW polarization modes. The Fourier components \( h_{\boldsymbol{k}}^s \) are defined via the transverse-traceless tensor decomposition:  
\begin{equation}\label{eq:hij}  
    h_{ij}(\eta, \boldsymbol{x}) = \int \frac{\mathrm{d}^3 \boldsymbol{k}}{(2\pi)^{3/2}} \left[ h_{\boldsymbol{k}}^{+}(\eta) e_{ij}^{+}(\boldsymbol{k}) + h_{\boldsymbol{k}}^{\times}(\eta) e_{ij}^{\times}(\boldsymbol{k}) \right] e^{i \boldsymbol{k} \cdot \boldsymbol{x}},  
\end{equation}  
with the transverse-traceless polarization tensors:  
\begin{eqnarray}  
    e_{ij}^{+}(\boldsymbol{k}) & = & \frac{1}{\sqrt{2}} \left[ e_i(\boldsymbol{k}) e_j(\boldsymbol{k}) - \bar{e}_i(\boldsymbol{k}) \bar{e}_j(\boldsymbol{k}) \right], \\  
    e_{ij}^{\times}(\boldsymbol{k}) & = & \frac{1}{\sqrt{2}} \left[ e_i(\boldsymbol{k}) \bar{e}_j(\boldsymbol{k}) + \bar{e}_i(\boldsymbol{k}) e_j(\boldsymbol{k}) \right],  
\end{eqnarray}  
where \( e(\boldsymbol{k}) \) and \( \bar{e}(\boldsymbol{k}) \) are orthonormal basis vectors spanning the plane perpendicular to the wavevector \( \boldsymbol{k} \).
The orthonormal vector triad \(\{e, \bar{e}, \boldsymbol{k}\}\), defining the spatial coordinate system, is illustrated in Fig.~\ref{fig:sc}.
The Fourier-space source term for GW generation is~\cite{Kohri:2018awv}:  
\begin{equation}\label{eq:source0}
    S_{\bsk}^s(\eta)=\int \frac{\mathrm{d}^3 \bsq}{(2 \pi)^{3 / 2}} e_{i j}^s(\bsk) q_i q_j
    \left[2 \Phi_{\bsq} \Phi_{\bsk-\bsq}+\frac{4}{3(1+\omega)}\left(\mathcal{H}^{-1} 
    \Phi_{\bsq}^{\prime}+\Phi_{\bsq}\right)\left(\mathcal{H}^{-1} 
    \Phi_{\bsk-\bsq}^{\prime}+\Phi_{\bsk-\bsq}\right)\right],
\end{equation}
where \(\omega = p/\rho\) is the universe’s equation-of-state parameter.  
Substituting the potential \(\Phi_{\boldsymbol{k}}\) from Eq.~\eqref{eq:Phidp}, the source term reduces to~\cite{Papanikolaou:2020qtd}:  
\begin{equation}\label{eq:S2}
    S_{\bsk}^s(\eta)=\int \frac{\mathrm{d}^3 \bsq}{(2 \pi)^{3 / 2}} e^s(\bsk, \bsq) 
    F(\bsq, \bsk-\bsq, \eta) \phi_{\bsq} \phi_{\bsk-\bsq},
\end{equation}
where \(e^s(\boldsymbol{k}, \boldsymbol{q}) \equiv e_{ij}^s(\boldsymbol{k}) q_i q_j\) encodes the polarization-projected momentum coupling.  
The kernel \(F\) is defined as:  
\begin{equation}\label{eq:F1}
    \begin{aligned}
    F(\bsq, \bsk-\bsq, \eta)&=2 T(q \eta) T(|\bsk-\bsq| \eta)+\frac{4}{3(1+\omega)}\left[\mathcal{H}^{-1} q T^{\prime}(q \eta)+T(q \eta)\right] \\
    &\times \left[\mathcal{H}^{-1}|\bsk-\bsq| T^{\prime}(|\bsk-\bsq| \eta)+T(|\bsk-\bsq| \eta)\right].
    \end{aligned}
\end{equation}
The solution to Eq.~\eqref{eq:hkEOM} is derived via the Green's function method:  
\begin{equation}\label{eq:greensolution}  
    h_{\boldsymbol{k}}^s(\eta) = \frac{4}{a(\eta)} \int_{\eta_i}^\eta \mathrm{d}\bar{\eta} \, G_{\boldsymbol{k}}(\eta, \bar{\eta}) a(\bar{\eta}) S_{\boldsymbol{k}}^s(\bar{\eta}),  
\end{equation}  
where the Green’s function \( G_{\boldsymbol{k}}(\eta, \bar{\eta}) \) satisfies:  
\begin{equation}\label{eq:G1}  
    G_{\boldsymbol{k}}^{\prime \prime}(\eta, \bar{\eta}) + \left[k^2 - \frac{a^{\prime \prime}(\eta)}{a(\eta)}\right] G_{\boldsymbol{k}}(\eta, \bar{\eta}) = \delta(\eta - \bar{\eta}),  
\end{equation}  
with initial conditions:  
\begin{equation}  
    \lim_{\eta \to \bar{\eta}} G_{\boldsymbol{k}}(\eta, \bar{\eta}) = 0 \quad \text{and} \quad \lim_{\eta \to \bar{\eta}} G_{\boldsymbol{k}}^{\prime}(\eta, \bar{\eta}) = 1.  
\end{equation} 
For a universe with scale factor \( a(\eta) \propto \eta^{2/(1+3\omega)} \), Eq.~\eqref{eq:G1} becomes:  
\begin{equation}  
    G_{\boldsymbol{k}}^{\prime \prime}(\eta, \bar{\eta}) + \left[k^2 - \frac{2(1-3\omega)}{(1+3\omega)\eta^2}\right] G_{\boldsymbol{k}}(\eta, \bar{\eta}) = \delta(\eta - \bar{\eta}).  
\end{equation}  
The general solution is:  
\begin{equation}  
    kG_{\boldsymbol{k}}(\eta, \bar{\eta}) = \frac{\pi}{2} \sqrt{x \bar{x}} \left[Y_n(x) J_n(\bar{x}) - J_n(x) Y_n(\bar{x})\right],  
\end{equation}  
where \( J_n(x) \) and \( Y_n(x) \) are Bessel functions of the first and second kind, respectively, with order \( n = \frac{1}{2}\sqrt{\frac{3(3-7\omega)}{1+3\omega}} \).
In a radiation-dominated universe (\( \omega = 1/3 \)):  
\begin{equation}\label{eq:RDUGreen}  
    kG_{\boldsymbol{k}}(\eta, \bar{\eta}) = \sin(x - \bar{x}).  
\end{equation}
Equations~\eqref{eq:S2}-\eqref{eq:greensolution} constitute the core formalism for determining tensor perturbations. The energy density of these tensor modes is derived in the following subsection.

\subsection{Tensor power spectrum}
GWs can be distinguished from the background within a length scale \(\lambda \ll \ell \ll L_B\), where \(\lambda\) is the wavelength of the GW and \(L_B\) represents the typical length scale of the background~\cite{Caprini:2018mtu}. In this context, the energy density of the GWs is expressed as follows:
\begin{equation}\label{eq:rhogw1}
    \rho_{\rm GW}(\eta,\boldsymbol{x})= 
    \sum_{s=+, \times} \frac{M_{\mathrm{Pl}}^2}{32 a^2} \overline{\left\langle \nabla_i h_{\alpha \beta}^s\nabla^i h_{\alpha \beta}^s \right\rangle},
\end{equation}
where \(M_{\rm Pl} = 1/\sqrt{8\pi G}\) is the reduced Planck mass, \(i\) denotes the spatial components, the overline indicates the oscillation average over the length scale \(\ell\), and the brackets represent an ensemble average. Utilizing the Fourier expansion of the GW given in Eq.~\eqref{eq:hij}, we have:
\begin{eqnarray}\label{eq:rhogw2}
    \rho_{\rm GW}(\eta,\boldsymbol{x})=\frac{M_{\mathrm{Pl}}^2}{32 a^2(2 \pi)^3} \sum_{s=+, \mathrm{x}} \int \mathrm{~d}^3 \bsk_1 
    \int \mathrm{~d}^3 \bsk_2 k_1 k_2 \overline{\left\langle h_{\bsk_1}^s(\eta) h_{\bsk_2}^{s, *}(\eta)
    \right\rangle} e^{i\left(\bsk_1-\bsk_2\right) \cdot \boldsymbol{x}}.
\end{eqnarray}
Let us begin by calculating the correlation function of the tensor as indicated in Eq.~\eqref{eq:rhogw2}. By inserting the Green's function solution given in Eq.~\eqref{eq:greensolution} for the GW, the correlation function can be expressed as:
\begin{equation}\label{eq:hh1}
    \left\langle h_{\bsk_1}^{s_1}(\eta) h_{\bsk_2}^{s_2, *}(\eta)\right\rangle=\frac{16}{a^2(\eta)} \int_{\eta_{i}}^\eta 
    \mathrm{d} \bar{\eta}_1 G_{k_1}\left(\eta, \bar{\eta}_1\right) a\left(\bar{\eta}_1\right) \int_{\eta_{i}}^\eta \mathrm{d} \bar{\eta}_2 
    G_{k_2}\left(\eta, \bar{\eta}_2\right) a\left(\bar{\eta}_2\right)\left\langle S_{\bsk_1}^{s_1}\left(\bar{\eta}_1\right) 
    S_{\bsk_2}^{s_2, *}\left(\bar{\eta}_2\right)\right\rangle,
\end{equation}
With the source term expressed in Eq.~\eqref{eq:S2}, we can write:
\begin{equation}\label{eq:sourcefunction}
    \begin{aligned}
    \left\langle S_{\bsk_1}^{s_1}\left(\bar{\eta}_1\right) S_{\bsk_2}^{s_2, *}\left(\bar{\eta}_2\right)\right\rangle
    &=\int \frac{\mathrm{d}^3 \bsq_1}{(2 \pi)^{3 / 2}} e^{s_1}\left(\bsk_1, \bsq_1\right) F\left(\bsq_1, \bsk_1-\bsq_1, \bar{\eta}_1\right) \\
    & \times \int \frac{\mathrm{d}^3 \bsq_2}{(2 \pi)^{3 / 2}} e^{s_2}\left(\bsk_2, \bsq_2\right) F^*\left(\bsq_2, \bsk_2-\bsq_2, \bar{\eta}_2\right)
    \left\langle\phi_{\bsq_1} \phi_{\bsk_1-\bsq_1} \phi_{\bsq_2}^* \phi_{\bsk_2-\bsq_2}^*\right\rangle .
    \end{aligned}
\end{equation}
Using the Wick theorem, the four-point correlation function can be decomposed as follows:
\begin{equation}\label{eq:Wickfp}
    \begin{aligned}
    &\left\langle\Phi_{\bsm_1}(\bar{\eta}_1) \Phi_{\bsm_2}(\bar{\eta}_1) \Phi_{\bsm_3}^*(\bar{\eta}_2) \Phi_{\bsm_4}^*(\bar{\eta}_2)\right\rangle=
    \left\langle\Phi_{\bsm_1}(\bar{\eta}_1) \Phi_{\bsm_2}(\bar{\eta}_1)\right\rangle \left\langle\Phi_{\bsm_3}^*(\bar{\eta}_2) \Phi_{\bsm_4}^*(\bar{\eta}_2)\right\rangle\\
    &+\left\langle\Phi_{\bsm_1}(\bar{\eta}_1) \Phi_{\bsm_3}^*(\bar{\eta}_2)\right\rangle \left\langle\Phi_{\bsm_2}(\bar{\eta}_1) \Phi_{\bsm_4}^*(\bar{\eta}_2)\right\rangle
    +\left\langle\Phi_{\bsm_1}(\bar{\eta}_1) \Phi_{\bsm_4}^*(\bar{\eta}_2)\right\rangle \left\langle\Phi_{\bsm_2}(\bar{\eta}_1) \Phi_{\bsm_3}^*(\bar{\eta}_2)\right\rangle\\
    &=\delta(\bsm_1-\bsm_2)\delta(\bsm_3-\bsm_4)n_{\Phi}(\bar{\eta}_1)n_{\Phi}^*(\bar{\eta}_2)
    +2\left\langle\Phi_{\bsm_1}(\bar{\eta}_1) \Phi_{\bsm_3}^*(\bar{\eta}_2)\right\rangle \left\langle\Phi_{\bsm_2}(\bar{\eta}_1) \Phi_{\bsm_4}^*(\bar{\eta}_2)\right\rangle,
    \end{aligned}
\end{equation}
where \( \boldsymbol{m}_1 = \boldsymbol{q}_1 \), \( \boldsymbol{m}_2 = \boldsymbol{k}_1 - \boldsymbol{q}_1 \), \( \boldsymbol{m}_3 = \boldsymbol{q}_2 \), and \( \boldsymbol{m}_4 = \boldsymbol{k}_2 - \boldsymbol{q}_2 \). The first term on the right-hand side contains Dirac delta functions \( \delta(\boldsymbol{m}_1 - \boldsymbol{m}_2)\delta(\boldsymbol{m}_3 - \boldsymbol{m}_4) \), where \( n_\Phi(\bar{\eta}) \) denotes the power spectrum.
We can simplify the expansion of the four-point correlation function by considering the following observations:
\begin{itemize}
    \item[1.] Symmetry and Momentum Integration: The integration detailed in Eq.~\eqref{eq:sourcefunction} exhibits invariance under the transformation of momentum transfers given by $\bsq_i\to \bsk_i-\bsq_i~(i=1,2)$. Consequently, both terms in the second line of Eq.~\eqref{eq:Wickfp} contribute equally to the source correlator \eqref{eq:sourcefunction}.   
    \item[2.] Vanishing Contribution: The presence of Dirac delta functions in the first term of the third line of Eq.~\eqref{eq:Wickfp} imposes the conditions $\bsq_1=\bsk_1/2$ and $\bsq_2=\bsk_2/2$. This situation leads to the conclusion that $\bsq$ is orthogonal to both $e(\bsk)$ and $\bar{e}(\bsk)$. Therefore, in this specific context, we ascertain that $e^s(\bsk,\bsq)\equiv 0$.
\end{itemize}
In light of these considerations, it is evident that the first term in the third line of Eq.~\eqref{eq:Wickfp} does not contribute to the source correlator \eqref{eq:sourcefunction}.

Utilizing the definition provided in Eq.~\eqref{eq:Phipw} for the power spectrum of scalar perturbations, we obtain
\begin{equation}
    \left\langle\phi_{\bsm_1}\phi_{\bsm_3}^*\right\rangle \left\langle\phi_{\bsm_2}\phi_{\bsm_4}^*\right\rangle
    =\delta(\bsk_1-\bsk_2)\delta(\bsq_1-\bsq_2)\left(\frac{2\pi^2}{q_1^3}\right)\left(\frac{2\pi^2}{(k_1-q_1)^3}\right)\mcp_{\phi}(q_1)\mcp_{\phi}(k_1-q_1).
\end{equation}
In the subsequent discussion, we adopt the following definitions: let $\bsk_1=\bsk_2\equiv \bsk$, $\bsq_1=\bsq_2\equiv \bsq$, $u=|\bsk-\bsq|/k$, and $v=q/k$. Under these definitions, Eq.~\eqref{eq:sourcefunction} can be expressed as
\begin{equation}\label{eq:source2}
    \begin{aligned}
    \left\langle S_{\bsk_1}^{s_1}\left(\bar{\eta}_1\right) S_{\bsk_2}^{s_2, *}\left(\bar{\eta}_2\right)\right\rangle 
    & =\frac{\pi}{2} \delta\left(\bsk_1-\bsk_2\right) \int \mathrm{d}^3 \bsq e^{s_1}\left(\bsk, \bsq\right) e^{s_2}\left(\bsk, \bsq\right) \\
    & \times F_k\left(u, v, \bar{\eta}_1\right) F_k^*\left(u, v, \bar{\eta}_2\right) \frac{\mcp_{\phi}\left(q\right)}{q^3} 
    \frac{\mcp_{\phi}\left(\left|\bsk-\bsq\right|\right)}{\left|\bsk-\bsq\right|^3},
\end{aligned}
\end{equation}
where \( F_k(u, v, \eta) \equiv F(\bsq, |\bsk - \bsq|, \eta) \). Utilizing the relations \( q = vk \) and \( \cos\theta = [k^2 + q^2 - (uk)^2]/(2kq) = (1 + v^2 - u^2)/(2v) \), we can transform the integration variables from \((q, \theta, \varphi)\) to \((v, u, \varphi)\). Specifically, this conversion can be expressed as follows:
\begin{equation}
    \int d^3\bsq=\int_{0}^{\infty}q^2\md q\int_{-1}^{1}\md (-\cos\theta)\int_{0}^{2\pi}\md \varphi
    =2k^3\int_{0}^{\infty}v\md v\int_{|1-v|}^{1+v}u\md u\int_{0}^{2\pi}\md\varphi.
\end{equation}
It is important to note that the functions \( F(u, v, \eta) \) and \( \mcp_{\phi}(k) \) are independent of the variable \( \varphi \). Consequently, we can directly integrate out the variable \( \varphi \) in Eq.~\eqref{eq:source2}. Referring to Fig.~\ref{fig:sc}, we have the following relations: \( e(\bsk) \cdot \bsq = q \sin \theta \cos \varphi \), \( \bar{e}(\bsk) \cdot \bsq = q \sin \theta \sin \varphi \), and \( e_{ij}^{s_1} e_{ij}^{s_2} = \delta^{s_1 s_2} \). Thus, the integration over \( \varphi \) can be expressed as follows:
\begin{eqnarray}
    \int_{0}^{2\pi} e^{s_1}\left(\bsk, \bsq\right) e^{s_2}\left(\bsk, \bsq\right)\mathrm{d}\varphi=\left\{\begin{matrix}
        \frac{\pi k^4}{32}\left [ 4v^2-(1-u^2+v^2)^2 \right ]^2,~~{\rm for}~~s_1=s_2, \\
        0,~~{\rm for}~~s_1\neq s_2 .
        \end{matrix}\right .
\end{eqnarray}
With these results, Eq.~\eqref{eq:source2} can be rewritten as follows:
\begin{eqnarray}\label{eq:source3}
    \begin{aligned}
    \left\langle S_{\bsk_1}^{s_1}\left(\bar{\eta}_1\right) S_{\bsk_2}^{s_2, *}\left(\bar{\eta}_2\right)\right\rangle 
    &=\frac{\pi^2}{2} \delta\left(\bsk_1-\bsk_2\right)\delta^{s_1s_2}\int_{0}^{\infty}\md v\int_{|1-v|}^{1+v}\md u\left [ \frac{4v^2-(1-u^2+v^2)^2}{4uv}\right ]^2\\
    &\times kF(u,v,\bar{\eta}_1)F^*(u,v,\bar{\eta}_2)\mcp_{\phi}(vk)\mcp_{\phi}(uk).
    \end{aligned}
\end{eqnarray}
We define the reduced GW power spectrum $\mathcal{P}_h\left(\eta, k\right)$ as:
\begin{equation}\label{eq:gwpwdef}
    \left\langle h_{\bsk_1}^{s_1}(\eta) h_{\bsk_2}^{s_2, *}(\eta)\right\rangle \equiv \delta^{(3)}\left(\bsk_1-\bsk_2\right) \delta^{s_1s_2} 
    \frac{2 \pi^2}{k^3} \mathcal{P}_h\left(\eta, k\right),
\end{equation}
where we set $\bsk_1 = \bsk_2 \equiv \bsk$. By combining these definitions with Eqs.~\eqref{eq:hh1} and~\eqref{eq:source3}, the GW power spectrum becomes:
\begin{equation}\label{eq:psp}
    \mathcal{P}_h(\eta, k)=4 \int_0^{\infty} \mathrm{d} v \int_{|1-v|}^{1+v} \mathrm{~d} u\left[\frac{4 v^2-\left(1-u^2+v^2\right)^2}{4uv}\right]^2 
    I^2(u,v,x) \mcp_{\phi}(vk) \mcp_{\phi}(uk),
\end{equation}
with the kernel function defined as:
\begin{equation}\label{eq:I}
    I(u, v, x)=\int_{x_g}^x \mathrm{d} \bar{x} \frac{a(\bar{x})}{a(x)} k G_k(x,\bar{x}) F_k(u,v,\bar{x}).
\end{equation}
Here, $x = k\eta$ and $x_g = k\eta_g$, where $\eta_g$ denotes the time of GW generation.

The fraction of the GW energy density per logarithmic wavelength, denoted as \(\Omega_{\rm GW}(\eta, k)\), is defined by:
\begin{equation}
    \rho_{\rm GW}(\eta )\equiv \rho_{\mathrm{tot}}(\eta) \int \Omega_{\mathrm{GW}}(\eta, k) \mathrm{d} \ln k,
\end{equation}
where \(\rho_{\rm tot}=3M_{\rm Pl}^2\mathcal{H}^2/a^2\) represents the total energy density of the universe. Utilizing Eqs.~\eqref{eq:rhogw2} and \eqref{eq:gwpwdef}, we obtain:
\begin{equation}
    \Omega_{\mathrm{GW}}(\eta, k)=\frac{1}{48}\left[ \frac{k}{\mathcal{H}(\eta)} \right]^2 \overline{\mathcal{P}_h(\eta, k)} 
    =\frac{1}{48}x^2\overline{\mathcal{P}_h(\eta, k)} .
\end{equation}
In the second equality, we have utilized the relation \(\mathcal{H}=1/\eta\) pertinent to the radiation-dominated universe. The spectrum of GW is directly influenced by its power spectrum. In this section, we have computed the general induced GWs resulting from the second-order scalar perturbations. Once the curvature perturbation and its power spectrum are established, we can calculate the correlator of the source and subsequently determine the power spectrum of the tensor modes.

\section{GW from perturbation of DW}\label{sec:GWfromDW}
With the results presented in sections~\ref{sec:perturb} and~\ref{sec:IGW}, we can now proceed to calculate the second-order tensor power spectrum resulting from the scalar perturbations of the DW network in a radiation-dominated universe.

\subsection{GW spectrum before DW annihilation}
Before the DW annihilation occurs at a conformal time \(\eta_a\), the DWs serve as a continuous source for the scalar perturbations, with the associated potential expressed as follows:
\begin{equation}\label{eq:phi2}
    \Phi_{k}(\eta)=\frac{Cf_w(\eta)}{2}\sqrt{\frac{\pi}{2k}}\delta\left(k-\frac{2\pi}{d_w}\right)~~{\rm for}~~\eta_i\le \eta\le \eta_a.
\end{equation}
Utilizing the transfer function presented in Eq.~\eqref{eq:Tphi} and noting that \(d(f(x)\delta(x-x_0))/dx=0\), we find that Eq.~\eqref{eq:F1} results in:
\begin{equation}
    F(u,v,x)=3T(vx)T(ux)=3(ux)^5(vx)^5\delta(u-v)\delta(ux-4\pi \ma).
\end{equation}
By applying Eq.~\eqref{eq:I} alongside the Green's function provided in Eq.~\eqref{eq:RDUGreen} for the radiation-dominated universe, we arrive at the following conclusion:
\begin{eqnarray}
    I_f(u, v, x) = 3\delta(u-v)u^4v^5\frac{\bar{x}^{11}\sin(x-\bar{x})}{x}\mid_{\bar{x}=\frac{4\pi \ma}{u}}.
\end{eqnarray}
Along with the power spectrum given in Eq.~\eqref{eq:mcp} for the potential, we obtain:
\begin{equation}
    \mathcal{P}_{h,f}(\eta, k)=\left [ \frac{3(4\pi \ma)^{11}B^2}{\pi^2k^8 x} \right ]^2\int_{{\rm max}(1/2,k_a/k)}^{k_i/k}
    \left( \frac{4v^2-1}{4v^2} \right)^2\left [ \frac{\sin\left(x- \frac{4\pi \ma}{v}\right)}{v^{10}}\right ]^2\md v,
\end{equation}
where \( k_i = 4\pi \ma/\eta_i \) and \( k_a = 4\pi \ma/\eta_a \) denote the comoving wavenumber at the times of formation and annihilation of the DWs, respectively. The condition \( v = u > |1 - v| \) necessitates that \( v > 1/2 \), and the momentum must lie within the range \( k_a \leq k \leq k_i \).
Our analysis focuses on the present-day GW spectrum, corresponding to the conformal time \(\eta \to \eta_0 = 1.47 \times 10^{18}~\text{s}\). In alignment with established methodology, we adopt the asymptotic \( x \to \infty \) limit to compute the contemporary GW spectrum, consistent with prior theoretical frameworks for cosmological gravitational wave predictions.
Averaging over the oscillations in the large time limit, the GW spectrum is found to be:
\begin{equation}\label{eq:ogf}
    \Omega_{\mathrm{GW},f}(\eta, k)=\frac{3}{32}\left [ \frac{(4\pi \ma)^{11}B^2}{\pi^2k^8} \right ]^2
    \left(-\frac{1}{42 s_a^{21}}+\frac{1}{368 s_a^{23}}+\frac{1}{19 s_a^{19}}-\frac{1}{19 s_b^{19}}+\frac{1}{42 s_b^{21}}-\frac{1}{368 s_b^{23}}\right),
\end{equation}
where we have employed the relations $k/\mathcal{H}=k\eta=x$, $s_a={\rm max}(1/2,k_a/k)$, and $s_b=k_i/k$.

In the following, we elucidate the physical significance of the late-time limit (\(x,\eta \to \infty\)) in the context of GW spectra. To anchor our analysis, we can adopt the DW annihilation time \(\eta_a\) as a reference and impose the condition \(\eta_0 / \eta_a \gg 1\), ensuring DW annihilation occurs in the early universe. For concreteness, if annihilation coincides with the QCD phase transition, \(\eta_a \simeq \eta_{\rm QCD} \simeq 10^6~\text{s}\). 

The late-time limit simplifies the dynamical evolution of GWs as follows:
\begin{itemize}
    \item[1.] Decoherence of transient dynamics: At late times, transient oscillations and source-dependent temporal variations in \(\mathcal{P}_h(\eta, k)\) average out, leaving only a time-independent amplitude.
    \item[2.] Scale-dependent amplification: The factor \(\left[\frac{k}{\mathcal{H}(\eta)}\right]^2\) governs the evolution of sub-horizon modes (\(k \gg \mathcal{H}\)), which dominate contributions to the spectrum. Conversely, super-horizon modes (\(k \ll \mathcal{H}\)) are suppressed due to this scaling.
    \begin{itemize}
        \item Note: In a radiation-dominated universe, \(\mathcal{H} \sim 1/\eta\), so \(k/\mathcal{H}\propto k\eta\). However, the decay of \(\mathcal{P}_h \propto \eta^{-2}\) (due to redshifting, see Eq.~\eqref{eq:hh1}) cancels the \((k\eta)^2\) growth, leaving \(\Omega_{\mathrm{GW}}\) constant at late times.
    \end{itemize}
    \item[3.] Freezing of gravitational waves: Sub-horizon GWs become ``frozen'' upon re-entering the Hubble horizon, with their present-day amplitude determined by this asymptotic regime.
\end{itemize}
The late-time limit in the calculation of the GW spectrum \(\Omega_{\mathrm{GW}}(\eta, k)\) corresponds to the asymptotic, steady-state behavior of the GW energy density after the source of the waves has ceased and the universe has expanded significantly.
This is critical for connecting theoretical predictions (e.g., from early-universe physics) to present-day observations by GW instruments.

\subsection{GW spectrum after DW annihilation}
\subsubsection{The evolution of the perturbation after annihilation}
After the annihilation of the DW at a time \(\eta_a\), the source is no longer present, and the evolution of the curvature perturbation is described by:
\begin{equation}\label{eq:phitime}
    \Phi_k^{\prime \prime}+3 \mathcal{H}\left(1+c_s^2\right) \Phi_k^{\prime}+c_s^2 k^2 \Phi_k=0~~{\rm for~~}\eta>\eta_a,
\end{equation}
where the speed of sound is given by \(c_s^2 = \omega\) and the expansion rate is expressed as $\mathcal{H}=a^{\prime}/a=2/((1+3\omega)\eta)$. Consequently, Eq.~\eqref{eq:phitime} can be transformed into a Bessel equation for \(\mathfrak{u}_k = \eta \Phi_k\).
\begin{equation}
    \mathfrak{u}_k^{\prime \prime}+\mu\frac{\mathfrak{u}_k^{\prime}}{\eta}+\left(\omega k^2-\frac{\mu}{\eta^2} \right)\mathfrak{u}_k=0,
\end{equation}
where $ \mu=\left[ \frac{6(1+\omega)}{1+3\omega}-2 \right]$. 
For a radiation-dominated universe, where \(c_s^2 \equiv \omega = 1/3\) and \(\mu = 2\), the solution is then given by:
\begin{equation}\label{eq:phi3}
    \Phi_k(\eta)=\frac{1}{z}\left[ C_1(k)j_1(z)+C_2(k)y_1(z) \right],
\end{equation}
where \( z = c_s k \eta \) (with \(\eta \geq \eta_a\)), and \( j_n(z) \) and \( y_n(z) \) are the spherical Bessel functions of order \( n \). To determine the constants \( C_1 \) and \( C_2 \), we require the continuity of both \(\Phi_k\) and \(\Phi_k^{\prime}\) at the time of DW annihilation, \(\eta_a\) (see also \cite{Domenech:2020ssp}). Taking the derivative of Eq.~\eqref{eq:phi3} with respect to the conformal time, we obtain:
\begin{equation}
    \Phi_k^{\prime}(\eta)=-\frac{c_sk}{z}\left[ C_1(k)j_2(z)+C_2(k)y_2(z) \right].
\end{equation}
Meanwhile, the derivative of Eq.~\eqref{eq:phi2} with respect to the conformal time is zero due to the presence of the Dirac delta function.
Using the relation $j_1(z)y_2(z)-y_1(z)j_2(z)=-1/z^2$, we obtain:
\begin{equation}
    C_1(k)=-z_a^3y_2(z_a)\Phi_k(\eta_a),{~~\rm and ~~}C_2(k)=z_a^3j_2(z_a)\Phi_k(\eta_a),
\end{equation}
where \( z_a = c_s k_a \eta_a = 4\pi \ma c_s \) as at the time of DW annihilation, \( k_a\equiv k(\eta_a)=4\pi \ma/\eta_a \). By utilizing Eq.~\eqref{eq:Tphi}, we then obtain:
\begin{eqnarray}
    C_1(k)&=&(4\pi \ma)^{8}c_s^3y_2(4\pi \ma c_s)\frac{B}{k^{5}\sqrt{k}},\\
    C_2(k)&=&(4\pi \ma)^{8}c_s^3j_2(4\pi \ma c_s)\frac{B}{k^{5}\sqrt{k}}.   
\end{eqnarray}
For \(\ma = 1\) and \(\omega = 1/3\), we find that \(j_2(z_a) = -0.139\) and \(y_2(z_a) = 0.026\). Consequently, we obtain the values \(C_1 = -10\Phi_k(\eta_a)\) and \(C_2 = -53\Phi_k(\eta_a)\).

\begin{figure}[t!]
    \centering
    \includegraphics[width=0.48\textwidth]{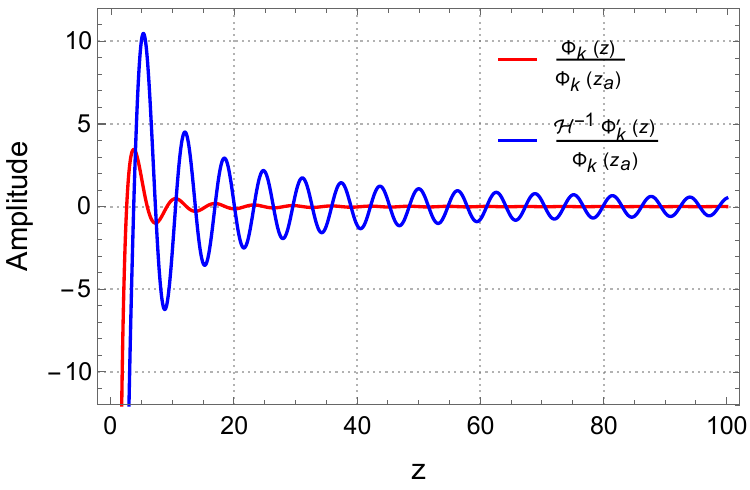}
    \includegraphics[width=0.48\textwidth]{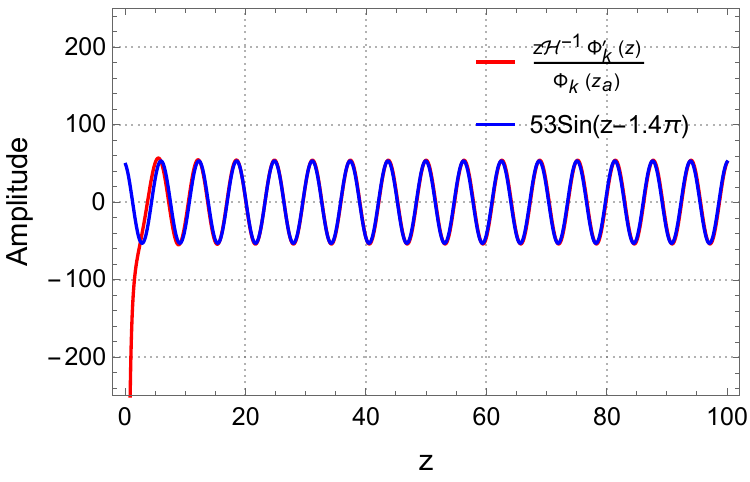}
    \caption{Left: The amplitudes of $\Phi_k(z)/\Phi_k(z_a)$ (red curve) and $\mathcal{H}^{-1}\Phi_k^{\prime}(z)/\Phi_k(z_a)$ (blue curve) are presented as a function of \(z\). Right: The amplitudes of $z\mathcal{H}^{-1}\Phi_k^{\prime}(z)/\Phi_k(z_a)$ (red curve) and $53\sin(z-1.4\pi)$ (blue curve) as a function of $z$. We assume that the annihilation of the DW occurs at a time \(\eta_a = 0\).}
     \label{fig:bessel}
\end{figure}
In the left panel of Fig.~\ref{fig:bessel}, we illustrate the amplitudes of $\Phi_k(z)/\Phi_k(z_a)$ and $\mathcal{H}^{-1}\Phi_k^{\prime}(z)/\Phi_k(z_a)$ as a function of \(z\). It is observed that, in comparison to \(\mathcal{H}^{-1}\Phi_k^{\prime}(z)\), \(\Phi_k(z)\) is negligible for the source term expressed in Eq.~\eqref{eq:source0}.
This implies that the time evolution of the gravitational potential dominates over the static scalar perturbation, thereby enhancing the production of scalar-induced GWs. This conclusion is valid for scalar perturbations in a radiation-dominated universe; however, it does not apply in a matter-dominated universe. In such a universe, the scalar perturbation remains constant, denoted as \(\Phi = \text{const}\) (this can be obtained by solving Eq.~\eqref{eq:phitime} with $c_s^2=\omega=0$ in a matter-dominated universe), leading to \(\Phi^{\prime} = 0\).
The disparity arises because \(\Phi\) oscillates due to radiation pressure, which results in a significant \(\Phi^{\prime}\), while radiation pressure suppresses \(\Phi\) on subhorizon scales. In contrast, non-relativistic matter is pressureless, meaning that \(\Phi\) remains constant throughout its evolution.
In the right panel of Fig.~\ref{fig:bessel}, we demonstrate that \(\mathcal{H}^{-1}\Phi_k^{\prime}(z)\) can be approximated by a sine function as \(z \to \infty\), specifically expressed as:
\begin{equation}
    \mathcal{H}^{-1}\Phi_k^{\prime}(z)\simeq -\frac{C_2(k)}{z}\sin(z-\theta_a),
\end{equation}
where \(\theta_a = 1.4\pi\). The exact value of \(\theta_a\) is not critical for our analysis. 
From Fig.~\ref{fig:bessel} we observe that the time derivatives of scalar perturbations \(\mathcal{H}^{-1}\Phi^{\prime}\) spike at the DW annihilation time \(z_a=0\), enhancing the source term for scalar-induced GWs. These observations indicate that the terms proportional to \(\mathcal{H}^{-1}\Phi_k^{\prime}(z)\) are the dominant components of the source term.
Consequently, the source term expressed in Eq.~\eqref{eq:source0} can be approximated as:
\begin{equation}
    S_{\bsk}^s(\eta)\simeq \int \frac{\mathrm{d}^3 \bsq}{(2 \pi)^{3 / 2}} e^s(\bsk, \bsq)
    \frac{C_2(vk)C_2(uk)}{uvz^2}\sin(vz-\theta_a)\sin(uz-\theta_a).
\end{equation}
Comparing with Eq.~\eqref{eq:S2} and the definition of the power spectrum given in Eq.~\eqref{eq:Phipw}, we have:
\begin{equation}\label{eq:FC2}
    F(u,v,z)=\frac{1}{uvz^2}\sin(vz-\theta_a)\sin(uz-\theta_a){~~\rm and~~}\mcp_{\phi}(k)=\frac{k^3C_2^2(k)}{2\pi^2},
\end{equation}
and 
\begin{equation}\label{eq:Iafter}
    I(u,v,x)=\int_{x_a}^{x}\md \bar{x}\frac{\bar{x}}{x}\sin(x-\bar{x})\frac{1}{uv\bar{z}^2}\sin(v\bar{z}-\theta_a)\sin(u\bar{z}-\theta_a).
\end{equation}
The integration for Eq.~\eqref{eq:Iafter} can be solved analytically using the following relations:
\begin{equation}\label{eq:intI}
    \begin{aligned}
    &\int_{x_a}^{x}\md \bar{x}\frac{1}{\bar{x}}\sin(x-\bar{x})\sin(a\bar{x}-\theta_a)\sin(b\bar{x}-\theta_a)=\\
    &\frac{1}{4} \left\{-\text{Ci}((a+b+1) x \right) \sin \left( 2 \theta _a+x \right) -\text{Ci}((a+b-1) x) \sin \left( x-2 \theta _a \right) +\text{Ci}((a+b+1) x_a)\sin \left(2 \theta _a+x\right)\\
    &+\text{Ci}((a+b-1) x_a) \sin \left(x-2 \theta _a\right)-\sin (x) \text{Ci}((a-b-1) x_a)-\sin (x) \text{Ci}((a-b+1)x_a)\\
    &+\sin (x)\text{Ci}((a-b-1) x)+\sin (x) \text{Ci}((a-b+1) x)-\text{Si}((a+b-1) x) \cos \left(x-2 \theta _a\right)\\
    &+\text{Si}((a+b+1) x)\cos \left(2 \theta _a+x\right)+\text{Si}((a+b-1) x_a) \cos \left(x-2 \theta _a\right)-\text{Si}((a+b+1) x_a) \cos \left(2 \theta _a+x\right)\\
    &-\cos (x)\text{Si}((a-b-1) x_a)+\cos (x) \text{Si}\left((a-b+1) x_a)+\cos (x) \text{Si}((a-b-1) x)-\cos (x) \text{Si}((a-b+1) x)\right\},
    \end{aligned}
\end{equation}
where $a=c_s v$ and $b=c_s u$. The functions $\text{Ci}(x)\equiv -\int_{x}^{\infty}\md y\cos(y)/y$ and $\text{Si}(x)\equiv \int_{0}^{x}\md y\sin(y)/y$ 
represent the cosine and sine integrals, respectively.

\subsubsection{Resonance and large $v$ contributions}

The results of Eq.~\eqref{eq:intI} can be approximated by two significant contributions: the resonant production of GWs and contributions from large \(v\) values~\cite{Inomata:2019ivs}.

{\bf Resonance:} The nonlinear nature of gravity in Einstein’s equations enables interactions between scalar perturbations with distinct oscillating modes. 
Such interactions can lead to resonance, a phenomenon where energy transfer or amplification occurs when the frequencies of interacting components satisfy conservation laws, such as energy and momentum matching.
In this context, resonance occurs when two slower modes nonlinearly interact to generate a faster third mode, requiring the faster mode’s frequency to equal the sum of the two slower frequencies. 
The faster propagation of the third mode is typically attributed to differences in dispersion relations, 
leading to a higher phase or group velocity compared to the original modes. 

In our scenario, the resonant production of GW arises when \(\text{Ci}(0) \to -\infty\) and \(\text{Si}(0) = 0\). Given the constraints \( |1-v| < u < 1+v \) alongside \( u > 0 \) and \( v > 0 \), we derive that \( u + v > 1 \) and \( u - v < 1 \). Consequently, resonant GW production occurs when the parameters satisfy the condition: \( u + v = 1/c_s = \sqrt{3} \).
Under this condition, the two perturbation modes induced by DWs can resonantly generate a faster-propagating mode in the plasma during wall annihilation.
The corresponding kernel function for the resonant production is given by:
\begin{equation}\label{eq:Ires}
    I_{\rm res}(u,v,x)=\frac{\sin (x-2\theta_a)}{4\omega uvx}\left\{\text{Ci}\left[ (c_s v+c_s u-1)x_a \right]-\text{Ci}\left[ (c_s v+c_s u-1)x \right]\right\},~{\rm with}~u+v=1/c_s.
\end{equation}

{\bf Large $v$ contribution:} 
The nonlinearity of Einstein’s equations implies that contributions from large \( v \) correspond to interactions between scalar modes with widely separated wavelengths — specifically, small-scale (high wavevector) modes coupling to large-scale (low wavevector) modes. Using the source term~\eqref{eq:source0}, we estimate  
\begin{equation}
    S_k\propto \int \md v v^4k^4\mathcal{F}(v,k,\eta),
\end{equation}
where \( \mathcal{F}(v, k, \eta) \) is a kernel function encoding the nonlinear coupling dynamics. The \( v^4 \) dependence indicates that large \( v \) values significantly amplify the source term, thereby dominating the overall contribution to \( S_k \) (except at the resonant scale).
The contributions in the limit of large \(u \simeq v \gg 1\) arise from the conditions \(\text{Si}(\infty) = \frac{\pi}{2}\) and \(\text{Ci}(\infty) = 0\). For late times, specifically as \(x \to \infty\), the kernel function in the large \(v\) limit is given by:
\begin{equation}
    I_{\rm LV}(u,v,x)=\frac{1}{4\omega uvx}\left[-\sin x\text{Ci}(x_a)+\cos x\left(\text{Si}(x_a)-\frac{\pi}{2}\right)\right]~~{\rm with}~~u=v\gg 1.
\end{equation}
The squared kernel functions averaged over the late time limit \(x \to \infty\) are expressed as follows~\cite{Inomata:2019ivs, Domenech:2020ssp}:
\begin{equation}\label{eqIres2}
    \overline{I_{\rm res}^2(u=c_s^{-1}-v,v,x_a)}\simeq \frac{1}{32\omega^2 u^2v^2x^2}\text{Ci}^2\left[ (c_s v+c_s u-1)x_a \right]
\end{equation}
and 
\begin{equation}\label{eq:ILV2}
    \overline{I_{\rm LV}^2(u=v,v,x)}\simeq \frac{1}{32\omega^2 u^2v^2x^2}\left[\text{Ci}^2(x_a)+\left(\text{Si}(x_a)-\frac{\pi}{2}\right)^2\right].
\end{equation}

\subsubsection{GW spectrum}
We now consider the GW spectrum arising from the perturbations following the annihilation of the DW. Depending on the different evolutionary stages of the DW, the kernel function for the perturbation can be separated into two parts:
\begin{equation}
    I(u,v,x)=I_f(u,v,x,x_i,x_a)+I_a(u,v,x,x_a),
\end{equation}
where \(I_f\) represents the kernel function prior to the annihilation of the wall, while \(I_a\) denotes the kernel function after the annihilation. It is reasonable to neglect the oscillation average of the cross term \(\overline{I_fI_a}\) in \(\overline{I^2}\) since it is generally understood that there is likely to be minimal correlation between modes created in different eras~\cite{Domenech:2020ssp}.
The energy density of the GW is then expressed as:
\begin{equation}
    \begin{aligned}
    \Omega_{\rm GW}(\eta,k)&=\frac{1}{48}\left(\frac{k}{\mathcal{H}} \right)^2\overline{\mcp_h(\eta,k)}
    \simeq \frac{1}{48}\left(\frac{k}{\mathcal{H}} \right)^2\left( \overline{\mcp_{h,f}(\eta,k)}+\overline{\mcp_{h,a}(\eta,k)} \right)\\
    &\equiv \Omega_{{\rm GW},f}(\eta,k)+\Omega_{{\rm GW},a}(\eta,k),        
    \end{aligned}
\end{equation}
with \(\Omega_{{\rm GW},f}(\eta,k)\) given by Eq.~\eqref{eq:ogf}, and \(\Omega_{{\rm GW},a}(\eta,k)\) representing the GW spectrum following the DW annihilation. Since the resonant production and large \(v\) contributions occur in two distinct parameter regimes, the power spectrum of the perturbation after the DW annihilation, \(\overline{\mcp_{h,a}(\eta,k)}\), can be approximated as:
\begin{equation}
    \overline{\mcp_{h,a}(\eta,k)}\simeq \overline{\mcp_{h,a,\rm res}(\eta,k)}+\overline{\mcp_{h,a, \rm LV}(\eta,k)}.
\end{equation}

Let us first calculate the resonant production of GW, which becomes significant within the parameter regime defined by the condition \( u + v \simeq c_s^{-1} \). Utilizing Equations~\eqref{eq:psp},~\eqref{eq:FC2}, and~\eqref{eqIres2}, we obtain:
\begin{equation}
    \begin{aligned}
    \overline{\mcp_{h,a,\rm res}(\eta,k)}&=\frac{1}{8\omega^2x^2}\int_{k_a/k}^{k_i/k}\md v\int_{{\rm max}(|1-v|,k_a/k)}^{{\rm min}(1+v,k_{i}/k)}\md u
    \left[\frac{4 v^2-\left(1-u^2+v^2\right)^2}{4u^2v^2}\right]^2 \\
    &\times \frac{v^3k^3C_2^2(vk)}{2\pi^2}\frac{u^3k^3C_2^2(uk)}{2\pi^2}\text{Ci}^2\left[ (c_s v+c_s u-1)x_a \right].       
    \end{aligned}
\end{equation}
To perform the integration, we change the integration variables to \( t = (c_s v + c_s u - 1) x_a \) and \( s = u - v \), with a Jacobian factor of \( 1/(2c_sx_a) \). The resonant contribution arises from the integration at the point \( t = 0 \). Therefore, we set \( t = 0 \) for the integrand, except for the \(\text{Ci}\) function~\cite{Inomata:2019ivs,Domenech:2020ssp}, which has a sharp value at this point. Integrating the \(\text{Ci}\) function yields a result approximately equal to \( \pi \); specifically, \(\int_{-\infty}^{\infty} \md t \, \text{Ci}^2(|t|) = \pi\). With these approximations, the integration is simplified as follows:
\begin{equation}
    \overline{\mcp_{h,a,\rm res}(\eta,k)}=\frac{4^5\omega^6(1-\omega)^2\widehat{C}_2^4}{\pi^3c_s x_a x^2k^{16}}\int_{s_1(k)}^{s_2(k)}\md s \frac{(1-s^2)^2}{(1-\omega s^2)^{10}},
\end{equation}
where \(\widehat{C}_2 = (4\pi \ma)^{8} c_s^3 j_2(4\pi \ma c_s) B\). 
The condition \(u > |1 - v|\) implies \(u > (c_s^{-1} - 1)/2\) (assuming \(c_s^{-1} > 1\)), which restricts the lower limit of integration to the range \(-1 \leq s_1(k) \leq 0\). 
Using the relation \(s = 2u - c_s^{-1}\), we find that \(u = c_s^{-1}/2\) when \(s_1(k) = 0\) and \(u = (c_s^{-1} - 1)/2\) when \(s_1(k) = -1\). 
Considering the limit \(u \geq u_{\text{min}} = k_a/k\), we have:
\begin{equation}
    s_1(k)=\left\{\begin{array}{lr}
    -1 & \qquad\frac{k_{a}}{k} < \frac{c_s^{-1}-1}{2}, \\
    2\frac{k_{a}}{k}-c_s^{-1} &  \qquad\frac{c_s^{-1}-1}{2}\leq \frac{k_{a}}{k} \leq  \frac{c_s^{-1}}{2}, \\
    0 & \qquad\frac{k_{a}}{k}>\frac{c_s^{-1}}{2}.
    \end{array}\right .
\end{equation}
Once again, the condition \( u < 1 + v \) limits the upper limit of the integration to the range \( 0 \leq s_0(k) \leq 1 \). We find that \( u = c_s^{-1}/2 \) when \( s_2(k) = 0 \), and \( u = (1 + c_s^{-1})/2 \) when \( s_2(k) = 1 \). Considering the limit \( u \leq u_{\rm max} = k_i/k \), we obtain:
\begin{equation}
    s_2(k)=\left\{\begin{array}{lr}
    1 & \qquad\frac{k_{i}}{k} > \frac{1+c_s^{-1}}{2}, \\
    2\frac{k_{i}}{k}-c_s^{-1} &  \qquad\frac{c_s^{-1}}{2}\leq \frac{k_{i}}{k} \leq  \frac{1+c_s^{-1}}{2}, \\
    0 & \qquad\frac{k_{i}}{k}<\frac{c_s^{-1}}{2}.
    \end{array}\right .
\end{equation}
With \( |s_{1(2)}| \sim 1 \), the result of the integration is approximately 2. We can also convert the integration variable \( s \) back to \( u \) using the relation \( s = 2u - c_s^{-1} \). Consequently, the resonant contribution to the GW spectrum is found to be
\begin{equation}\label{eq:GWres}
    \Omega_{{\rm GW},a,{\rm res}}(\eta,k)=\frac{4^5\omega^6(1-\omega)^2\widehat{C}_2^4}{48\pi^3c_s x_a k^{16}}\int_{u_a(k)}^{u_b(k)}\md u 
    \frac{2[1-(2u-c_s^{-1})^2]^2}{[1-\omega (2u-c_s^{-1})^2]^{10}},
\end{equation}
where $u_a(k)={\rm max}((c_s^{-1}-1)/2,k_a/k)$ and $u_b(k)={\rm min}((c_s^{-1}+1)/2,k_i/k)$.
The integration has an analytical expression, which is tedious and not shown here.

With the squared kernel function~\eqref{eq:ILV2}, the power spectrum arising from large $v$ contributions is given by 
\begin{equation}\label{eq:mcpLV1}
    \begin{aligned}
    \overline{\mcp_{h,a,\rm LV}(\eta,k)}&=\frac{1}{8\omega^2x^2}\int_{{\rm max}{(1/2,k_a/k)}}^{k_i/k}\md v\int_{{\rm max}(|1-v|,k_a/k)}^{{\rm min}(1+v,k_{i}/k)}\md u
    \left[\frac{4 v^2-\left(1-u^2+v^2\right)^2}{4u^2v^2}\right]^2 \\
    &\times \frac{v^3k^3C_2^2(vk)}{2\pi^2}\frac{u^3k^3C_2^2(uk)}{2\pi^2}\left[\text{Ci}^2(x_a)+\left(\text{Si}(x_a)-\frac{\pi}{2}\right)^2\right].       
    \end{aligned}
\end{equation}
For the significant contribution at large \( v \), we can approximate \( u \simeq v \gg 1 \). Thus, we can incorporate \( \delta(u-v) \) into Eq.~\eqref{eq:mcpLV1}, leading to the GW spectrum
\begin{equation}\label{eq:GWLV}
    \begin{aligned}
    \Omega_{{\rm GW},a,\rm LV}(\eta,k)&=\frac{\widehat{C}_2^4 }{1536\pi^4\omega^2k^{16}}\left[\text{Ci}^2(x_a)+\left(\text{Si}(x_a)-\frac{\pi}{2}\right)^2\right] \\
    &\times \left(-\frac{1}{42 s_a^{21}}+\frac{1}{368 s_a^{23}}+\frac{1}{19 s_a^{19}}-\frac{1}{19 s_b^{19}}+\frac{1}{42 s_b^{21}}-\frac{1}{368 s_b^{23}}\right),
    \end{aligned}
\end{equation}
where $s_a={\rm max}(1/2,k_a/k)$ and $s_b=k_i/k$. 
We observe that the $k$-dependence in Eq.~\eqref{eq:GWLV} is consistent with that in Eq.~\eqref{eq:ogf}; therefore, the GW spectra of both are identical apart from their amplitudes. This similarity arises because both spectra are derived under the condition $u=v$.

\subsection{Results}
\begin{figure}[t!]
    \centering
    \includegraphics[width=0.75\textwidth]{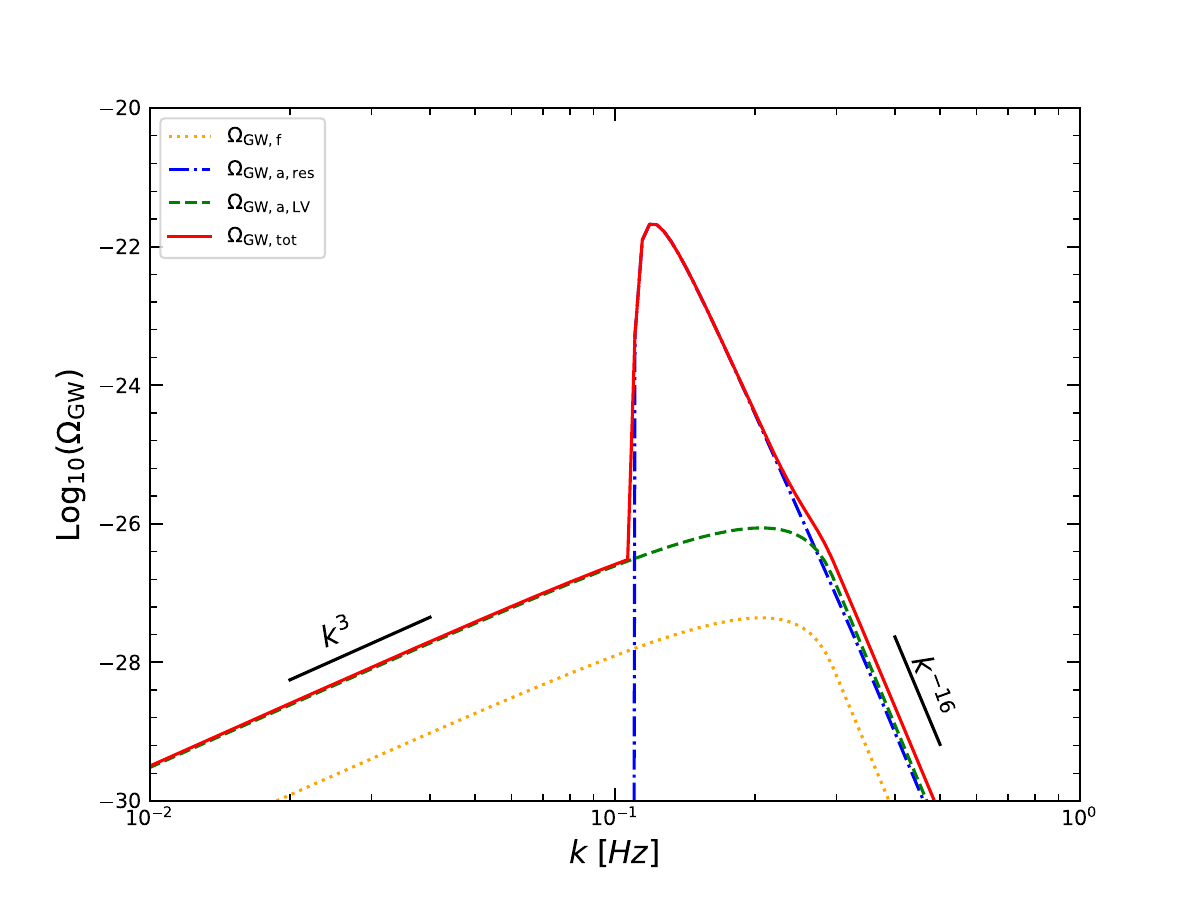}
    \caption{The induced GW spectrum \(\Omega_{\rm GW}\) from scalar perturbations associated with DWs is analyzed as a function of frequency today. In this context, we fix the formation temperature at \(T_f=10^8\)~GeV and the annihilation temperature at \(T_{\rm ann}=10^4\)~GeV. 
    In the resulting graph, the dotted curve represents the induced GW spectrum from the DW network prior to the annihilation of the domain walls. The blue dash-dotted curve corresponds to the resonant production of GWs, while the green dashed curve illustrates the contributions from large velocity effects. Lastly, the red solid curve depicts the total spectrum of induced gravitational waves.}
     \label{fig:ogw}
\end{figure}

The principal findings of this study are encapsulated in the GW spectrum, as delineated by Eqs.~\eqref{eq:ogf}, \eqref{eq:GWres}, and~\eqref{eq:GWLV}. The total GW spectrum resulting from the DW perturbations is expressed as the aggregate of these contributions:
\begin{equation}
    \Omega_{\rm GW,tot}=\Omega_{{\rm GW},f}+\Omega_{{\rm GW},a,{\rm res}}+\Omega_{{\rm GW},a,{\rm LV}}.
\end{equation}
The results are presented in Fig.~\ref{fig:ogw}. This figure assumes that DWs are generated at a cosmic temperature of \(T_f = 10^8\) GeV and subsequently annihilate at \(T_{\rm ann} = 10^4\) GeV, which corresponds to frequencies of \(k_i = 1492.3\) Hz and \(k_a = 0.15\) Hz for the scalar perturbation at the times of formation and annihilation, respectively. It is important to note that during this period, DWs do not exert a predominant influence on the universe.

In examining Fig.~\ref{fig:ogw}, one can observe that the shape of the GW spectrum from the DW prior to annihilation, indicated by the orange dotted line, resembles that of the large velocity contribution following DW annihilation, represented by the green dashed line. However, the amplitude of the former is substantially smaller than that of the latter, rendering it negligible. Resonant production of GWs, denoted by the blue dot-dashed line, occurs when the frequency of the GW closely matches that of the perturbation at annihilation, i.e., \(k\simeq k_a\). 
The spectrum of resonant production exhibits a peak at \(k\simeq k_a\), resulting in an enhancement of the GW signal by approximately four orders of magnitude.
For \(k\lesssim k_a\), the contribution from large $v$ dominates the spectrum.

It is noteworthy that the scalar perturbations arising from DWs are horizon-sized, which implies that the oscillation modes of the GWs are superhorizon if their frequencies are lower than that of the perturbation at the time of wall annihilation, i.e., \(k\lesssim k_a\). Consequently, causality dictates that the superhorizon modes should scale as \(k^3\), as illustrated in Fig.~\ref{fig:ogw}. In contrast, for higher frequencies \(k\gtrsim k_a\), the modes are within the horizon, leading to decay of the GWs at a rate of \(k^{-16}\).

Finally, the example depicted in Fig.~\ref{fig:ogw} indicates that the peak amplitude of the induced GW signal is approximately \(\Omega_{\rm GW}\sim 10^{-22}\), which exceeds the detection capabilities of current and near-future GW experiments. The power spectrum of the curvature perturbation associated with the DW is proportional to \(k^{-8}\), suggesting that the GW signals could be significantly enhanced and enter the detection range of experiments if the annihilation of DWs occurs at a considerably later time.

\begin{figure}[t!]
    \centering
    \includegraphics[width=0.49\textwidth]{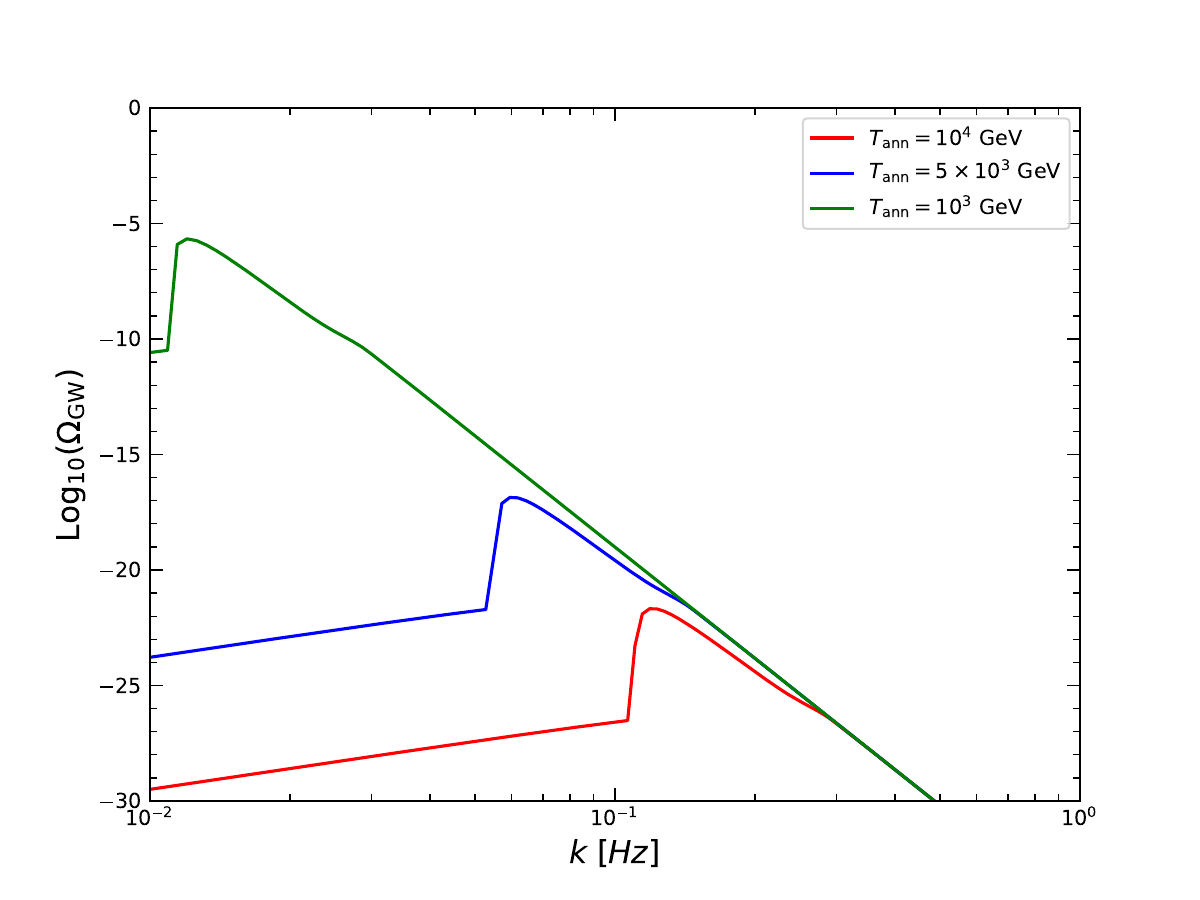}
    \includegraphics[width=0.49\textwidth]{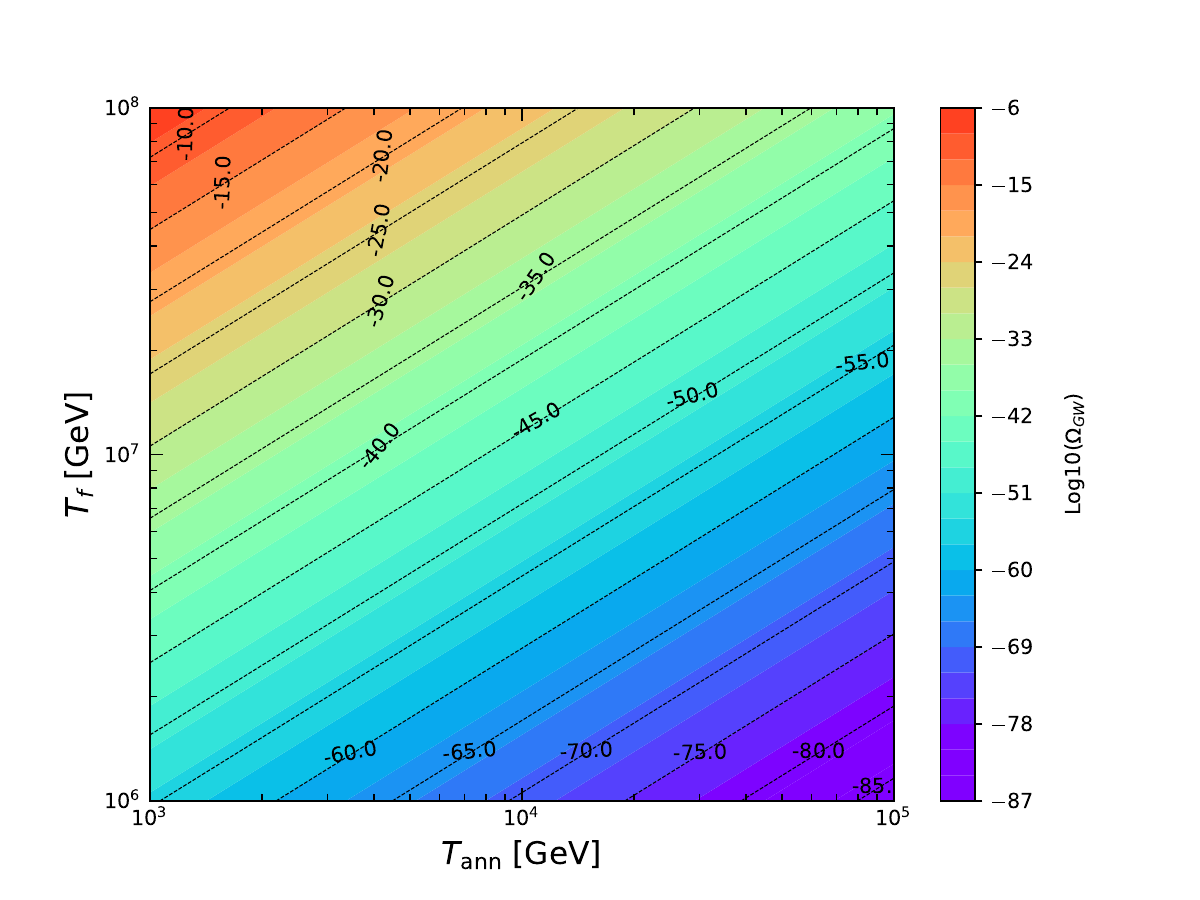}
    \caption{Left: The induced GW spectrum \(\Omega_{\rm GW}\) from DW scalar perturbations is depicted as a function of frequency today. In this analysis, we fix the formation temperature at \(T_f=10^8\)~GeV. The red, blue, and green curves represent the results for annihilation temperatures of \(T_{\rm ann}=10^{4}\)~GeV, \(5\times 10^{3}\)~GeV, and \(10^{3}\)~GeV, respectively.
    Right: The contour plot illustrates the peak amplitude of the GW spectrum as a function of both the formation and annihilation temperatures of the DW network.}
     \label{fig:temp}
\end{figure}

To elucidate this point further, the left panel of Fig.~\ref{fig:temp} presents the induced GW spectrum for three values of \(T_{\rm ann}\), while maintaining the formation temperature of the DW at \(T_f=10^8\) GeV. The amplitude of the GW scales as \(\Omega_{\rm GW}\propto T_{\rm ann}^{-16}\) (this relationship follows from \(\Omega_{\rm GW}\propto k_a^{-16}\propto \eta_a^{16}\)). Observations indicate that the peak amplitude of \(\Omega_{\rm GW}\) sharply increases from a negligible value of approximately \(\sim 10^{-22}\) at \(T_{\rm ann}=10^4\) GeV to an observable magnitude of about \(\sim 10^{-6}\) at \(T_{\rm ann}=10^3\) GeV. Furthermore, the peak frequency \(k_{\rm peak}\propto T_{\rm ann}\). The right panel of Fig.~\ref{fig:temp} depicts the contour of peak amplitude as a function of the formation and annihilation temperatures of the DW network. This figure demonstrates that \(\Omega_{\rm GW}\propto T_f^{24}\), as indicated by Eq.~\eqref{eq:GWres}, where it becomes evident that \(\Omega_{\rm GW}\propto \sigma_w^{8}\simeq T_{f}^{24}\), utilizing the estimation \(\sigma_w=T_f^3\).

In the prior literature, the GWs emitted during the annihilation of DWs have been extensively studied. The GW spectrum resulting from DW annihilation can be approximated by the following expression:
\begin{equation}\label{eq:gwspec}
    \Omega_{\rm{GW}}=\Omega_{\rm{GW}}^{\rm peak}\frac{(\gamma_c+c_1)^{c_2}}{(c_1x^{-\gamma_c/c_2}+\gamma_c x^{c_1/c_2})^{c_2}},
\end{equation}
where \(\Omega_{\rm GW}^{\rm peak}\) denotes the peak amplitude of the GW spectrum at the peak frequency \(k_{\rm peak}\) (refer to Ref.~\cite{Lu:2024ngi} for additional details), \(x=k/k_{\rm peak}\), and \(\gamma_c=3\) is introduced based on causality requirements~\cite{Cai:2019cdl}. The parameters \(c_1\) and \(c_2\) are taken to be approximately 1, as inferred from simulations; it should be noted that \(c_1\) and \(c_2\) may vary depending on the specific physical model~\cite{Hiramatsu:2012sc}.

\begin{figure}[t!]
    \centering
    \includegraphics[width=0.75\textwidth]{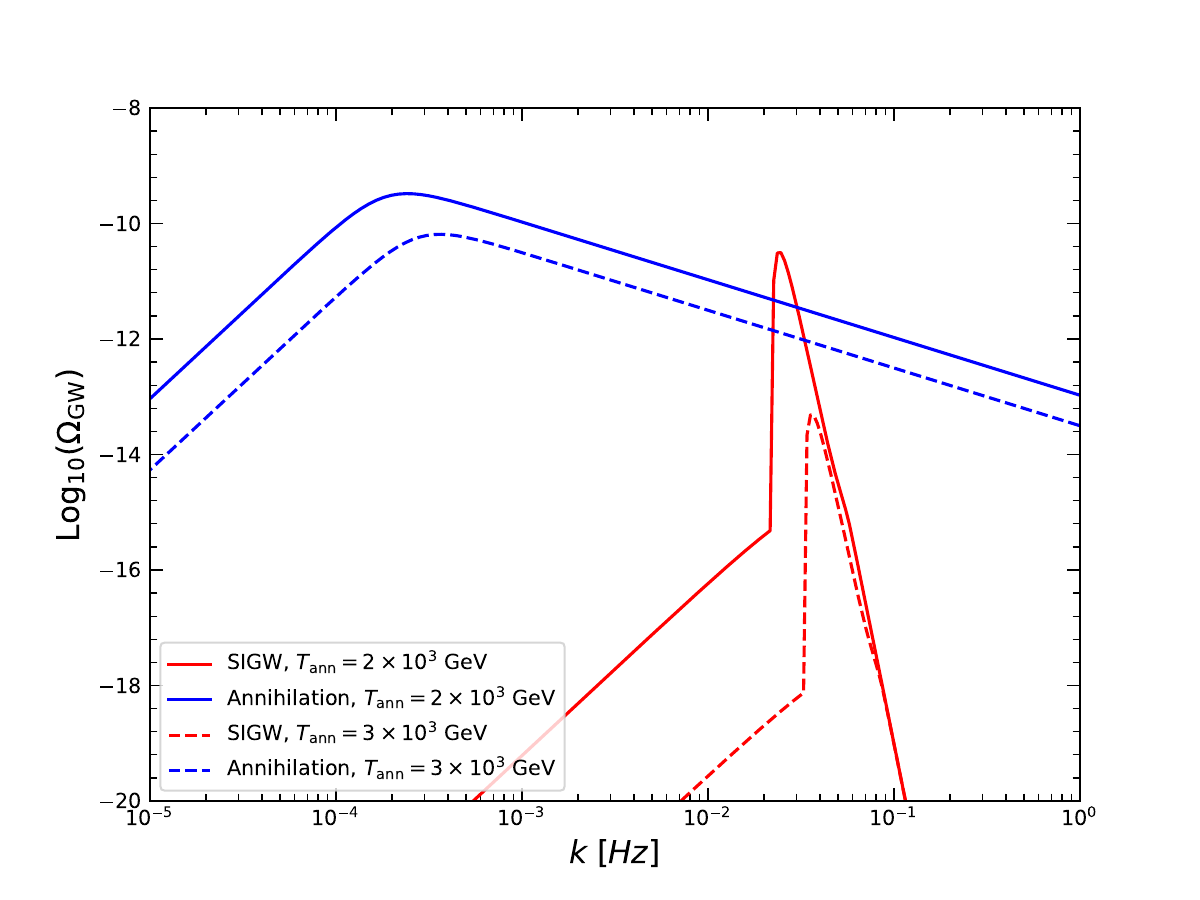}
    \caption{The GW spectrum \(\Omega_{\rm GW}\) is presented as a function of frequency today. In this plot, the blue curves represent the spectrum originating from the annihilation of the DW, while the red curves illustrate the induced GW resulting from perturbations within the DW network. The formation temperature is set at \(T_f=10^8\)~GeV. The solid curves correspond to the annihilation temperatures \(T_{\rm ann}=2\times 10^{3}\)~GeV, while the dashed curves correspond to \(T_{\rm ann}=3\times 10^{3}\)~GeV.}
     \label{fig:annisigw}
\end{figure}

In Fig.~\ref{fig:annisigw}, we present a comparison between the scalar-induced GW resulting from perturbations attributed to the DW network and the GW spectrum generated by the annihilation of the wall. In this figure, the formation temperature is fixed at \(T_f=10^8\)~GeV. The blue and red curves illustrate the spectra from the annihilation of the DW and the induced GW due to the perturbations from the DW network, respectively. We consider values of \(T_{\rm ann}=2\times 10^{3}\)~GeV and \(3\times 10^{3}\)~GeV for the solid and dashed curves, respectively.

Our analysis indicates that the amplitude of the scalar-induced GWs from the DW network becomes comparable to the GW signal from the annihilation of the DW when the annihilation occurs later, specifically at the temperature \(T_{\rm ann}=2\times 10^3\)~GeV.  
Using Eq.~\eqref{eq:Tdom}, the dominant temperature of the DW is calculated to be \(T_{\rm dom}\simeq 283\)~GeV for \(\sigma_w^{1/3}=T_f=10^8\)~GeV. Therefore, DWs generated at \(T_f=10^8\)~GeV do not dominate the universe at temperatures exceeding \(T\gtrsim 10^3\)~GeV.

In conclusion, we affirm that the scalar-induced GW signal from the DW network can indeed be comparable to the GW signal from the annihilation of the DW and may be detectable in future GW experiments, provided that the annihilation of the DW occurs at a sufficiently late time.

\section{Conclusions}\label{sec:conclusions}
The DW is a significant topological defect that arises during the spontaneous breaking of a discrete symmetry. Discrete symmetries are prevalent in various models of new particle physics; therefore, detecting the GW spectrum resulting from the evolution of DWs can serve as a unique method for probing physics beyond the Standard Model.

Previous literature has extensively examined the GWs emitted during the annihilation of DWs. In this work, we present, for the first time, the scalar-induced GW spectrum resulting from perturbations of the wall. Our findings indicate that the spectrum exhibits a pronounced peak at the frequency \(k \simeq k_a\) (where \(k_a\) represents the frequency of perturbation at the time of DW annihilation), significantly enhancing the signal to levels that may be detectable in experiments.
For subhorizon modes with \(k \gtrsim k_a\), the spectrum decays as \(\propto k^{-16}\). In contrast, for superhorizon modes with \(k \lesssim k_a\), the spectrum scales as \(\propto k^{3}\), which aligns with causality principles. 

The induced GW signals from DW perturbations can be distinguished from other GW sources, including first-order phase transitions, inflation, and cosmic strings. We will discuss the potential for detecting this signal across various GW experiments, including PTAs, LISA, Taiji, Tianqin, and LIGO, in forthcoming work.

\section*{Acknowledgements}
BQL is supported in part by the National Natural Science Foundation of China under Grant No.~12405058 and by the Zhejiang 
Provincial Natural Science Foundation of China under Grant No.~LQ23A050002.

\appendix

\section{Separation length}\label{app:splength}
The boundary at which an object's gravitational potential becomes position-dependent defines its characteristic length. For instance, consider a sphere with a uniform mass distribution and radius \(R\). The gravitational potential for this sphere can be approximated as \(U \sim -\frac{GM}{R}\) for \(r \lesssim R\) and \(U \sim -\frac{GM}{r}\) for \(r \gtrsim R\). Consequently, \(R\) is identified as the characteristic length of the sphere.
In the case of DWs, utilizing Equation \eqref{eq:pot1}, we obtain the following expressions for the gravitational potential:
\begin{eqnarray}
\tilde{\Psi} \sim \left\{
\begin{matrix}
2\pi G\sigma_w a d_w & {~~\rm for~~}x \lesssim d_w, \\
\pi G\sigma_w a \frac{d_w^2}{|x|} & {~~\rm for~~}x \gtrsim d_w.
\end{matrix}
\right.
\end{eqnarray}
From this, it is evident that \(d_w\) represents the characteristic length (in comoving coordinates) of the wall.

For a configuration of DWs distributed within space, the mean separation length between walls is determined by this characteristic length. This can be elucidated through the following analysis. The walls experience repulsion due to their gravitational effects and ultimately stabilize at positions corresponding to the extrema of the gravitational potential. Considering the potential described by Equation \eqref{eq:pot2}:
\begin{equation}
    \tilde{\Psi} = C \cos^2 \left( \frac{\pi x}{d_w} \right),
\end{equation}
for a network consisting of \(N\) walls, we can find the location of the extrema by applying the condition \(\frac{d\tilde{\Psi}}{dx} = 0\). This condition yields \(x = n d_w\) for \(n = 0, 1, \ldots, N-1\). Therefore, the mean separation length between walls in the network is characterized by the characteristic length of the wall.

We now demonstrate that the mean separation length \(L_w = a d_w\) can also be identified as the correlation length of the wall within the equations of motion described by Equation \eqref{eq:VOS}. Consider a configuration where \(N\) static walls are distributed along the \(x\) direction (noting that we now use \(x\) to represent the physical coordinate rather than the comoving coordinate). The average energy-momentum tensor within the mean separation length \(L_w\) (which corresponds to the characteristic length of the wall) can be expressed as follows:
\begin{eqnarray}
    \begin{aligned}
    \overline{T}_{\mu\nu}= \frac{1}{L_w}\int_{L_w} dx \, T_{\mu\nu}(x)
    =\frac{S_{\mu\nu}^x}{L_w},
    \end{aligned}   
\end{eqnarray}
where the surface energy-momentum tensor is defined as
\begin{equation}
    S_{\mu\nu}^x=\int_{L_w} dx \, T_{\mu\nu}(x),
\end{equation}
with the superscript \(x\) indicating the axis along which the wall is located. 

In general, DWs may be oriented in any arbitrary direction within three-dimensional space. The three-dimensional surface energy-momentum tensor in the wall frame is expressed as follows:
\begin{equation}\label{eq:Smunu}
    S_{\mu\nu}=\frac{\sigma_w}{3}\text{diag}(3,-2,-2,-2).
\end{equation}
For detailed calculations regarding \(S_{\mu\nu}\) with an arbitrary velocity, refer to Ref.~\cite{Lu:2024ngi}. 
Consequently, the mean wall energy density can be determined as:
\begin{equation}
    \rho_w = \overline{T}_{00} = \frac{S_{00}}{L_w} = \frac{\sigma_w}{L_w}.
\end{equation}
This leads us to obtain the one-scale assumption utilized in the equations of motion as presented in Equation~\eqref{eq:VOS}, which arises from the relationship between the mean separation length of the network and the wall energy density. Therefore, it is concluded that the mean separation length serves as the correlation length of the wall within the framework established by the equations of motion outlined in Equation~\eqref{eq:VOS}.

\section{DW evolution}\label{app:DWevolution}

The velocity-dependent one-scale (VOS) model serves as the canonical framework for the quantitative investigation of the evolution of defect networks. This model employs the so-called one-scale assumption to simplify the dynamics of defects, as outlined in the literature~\cite{Martins:2016book}. For a wall defect, the curvature of the wall surface is characterized by two curvature radii, denoted as \(R_1\) and \(R_2\). The one-scale assumption posits that these curvature radii share the same average value and are set equal to the correlation length, such that \(R_1 = R_2 = L_w\)~\cite{Martins:2016ois}. 
Under this assumption, the energy density of the wall can be expressed as
\begin{equation}\label{eq:onescale}
    \rho_{w}=\frac{\sigma_w}{L_w},
\end{equation}
where $L_w$ represents the wall's correlation length.
As demonstrated in appendix~\ref{app:splength}, the correlation length is indeed the mean separation length or characteristic length of the wall.
Using Eq.~\eqref{eq:onescale}, the evolution of the DW is described by the following equations (for further details, see~\cite{Martins:2016book}):
\begin{equation}\label{eq:VOS}
    \begin{aligned}
    \frac{d L_w}{d t} & = H L_w + \frac{L_w}{\ell_d} \beta v_w^2 + c_w \beta v_w, \\
    \frac{d \beta v_w}{d t} & = \left(1 - \beta v_w^2\right) \left(\frac{k_w}{L_w} - \frac{\beta v_w}{\ell_d}\right),
    \end{aligned}
\end{equation}
where \( v_w \) represents the averaged velocity of the wall. 
The momentum parameter \(k_w = 0.66 \pm 0.04\) and the chopping parameter \(c_w = 0.81 \pm 0.04\) are derived from simulations of DW evolution in a radiation-dominated universe~\cite{Martins:2016ois}. The momentum parameter reflects the curvature acceleration effect, which is significant for small wall sizes at the initial time. The chopping effect arises when DWs are subjected to forces causing them to move, collide, or interact with other walls, potentially leading to the fragmentation of larger domains into smaller ones. This chopping effect may be significantly reduced when closed configurations are formed.
The damping length is defined as 
\begin{equation}
    \frac{1}{\ell_d} = 3 H + \frac{1}{\ell_f},
\end{equation}
which accounts for the damping effects on the network arising from Hubble drag and particle friction. This particle friction is related to the particle pressure \(\Delta P\) as follows:
\begin{equation}
    \frac{1}{\ell_f} = \frac{\Delta P}{\beta v_w \sigma_w}.
\end{equation}
The friction can impede the expansion of the wall. In typical models, the effects of particle friction are often considered negligible.

\bibliographystyle{JHEP}
\bibliography{reference}

\providecommand{\href}[2]{#2}\begingroup\raggedright\begin{thebibliography}{10}

\bibitem{LIGOScientific:2016aoc}
{\bf LIGO Scientific, Virgo} Collaboration, B.~P. Abbott {\em et~al.}, {\it
  {Observation of Gravitational Waves from a Binary Black Hole Merger}},  {\em
  Phys. Rev. Lett.} {\bf 116} (2016), no.~6 061102,
  [\href{http://arxiv.org/abs/1602.03837}{{\tt 1602.03837}}].

\bibitem{NANOGrav:2023gor}
{\bf NANOGrav} Collaboration, G.~Agazie {\em et~al.}, {\it {The NANOGrav 15 yr
  Data Set: Evidence for a Gravitational-wave Background}},  {\em Astrophys. J.
  Lett.} {\bf 951} (2023), no.~1 L8,
  [\href{http://arxiv.org/abs/2306.16213}{{\tt 2306.16213}}].

\bibitem{EPTA:2023fyk}
{\bf EPTA} Collaboration, J.~Antoniadis {\em et~al.}, {\it {The second data
  release from the European Pulsar Timing Array III. Search for gravitational
  wave signals}},  \href{http://arxiv.org/abs/2306.16214}{{\tt 2306.16214}}.

\bibitem{Reardon:2023gzh}
D.~J. Reardon {\em et~al.}, {\it {Search for an Isotropic Gravitational-wave
  Background with the Parkes Pulsar Timing Array}},  {\em Astrophys. J. Lett.}
  {\bf 951} (2023), no.~1 L6, [\href{http://arxiv.org/abs/2306.16215}{{\tt
  2306.16215}}].

\bibitem{Xu:2023wog}
H.~Xu {\em et~al.}, {\it {Searching for the Nano-Hertz Stochastic Gravitational
  Wave Background with the Chinese Pulsar Timing Array Data Release I}},  {\em
  Res. Astron. Astrophys.} {\bf 23} (2023), no.~7 075024,
  [\href{http://arxiv.org/abs/2306.16216}{{\tt 2306.16216}}].

\bibitem{Caprini:2018mtu}
C.~Caprini and D.~G. Figueroa, {\it {Cosmological Backgrounds of Gravitational
  Waves}},  {\em Class. Quant. Grav.} {\bf 35} (2018), no.~16 163001,
  [\href{http://arxiv.org/abs/1801.04268}{{\tt 1801.04268}}].

\bibitem{Domenech:2021ztg}
G.~Dom\`enech, {\it {Scalar Induced Gravitational Waves Review}},  {\em
  Universe} {\bf 7} (2021), no.~11 398,
  [\href{http://arxiv.org/abs/2109.01398}{{\tt 2109.01398}}].

\bibitem{Planck:2018nkj}
{\bf Planck} Collaboration, N.~Aghanim {\em et~al.}, {\it {Planck 2018 results.
  I. Overview and the cosmological legacy of Planck}},  {\em Astron.
  Astrophys.} {\bf 641} (2020) A1, [\href{http://arxiv.org/abs/1807.06205}{{\tt
  1807.06205}}].

\bibitem{Byrnes:2018txb}
C.~T. Byrnes, P.~S. Cole, and S.~P. Patil, {\it {Steepest growth of the power
  spectrum and primordial black holes}},  {\em JCAP} {\bf 06} (2019) 028,
  [\href{http://arxiv.org/abs/1811.11158}{{\tt 1811.11158}}].

\bibitem{Ananda:2006af}
K.~N. Ananda, C.~Clarkson, and D.~Wands, {\it {The Cosmological gravitational
  wave background from primordial density perturbations}},  {\em Phys. Rev. D}
  {\bf 75} (2007) 123518, [\href{http://arxiv.org/abs/gr-qc/0612013}{{\tt
  gr-qc/0612013}}].

\bibitem{Bugaev:2009zh}
E.~Bugaev and P.~Klimai, {\it {Induced gravitational wave background and
  primordial black holes}},  {\em Phys. Rev. D} {\bf 81} (2010) 023517,
  [\href{http://arxiv.org/abs/0908.0664}{{\tt 0908.0664}}].

\bibitem{Alabidi:2012ex}
L.~Alabidi, K.~Kohri, M.~Sasaki, and Y.~Sendouda, {\it {Observable Spectra of
  Induced Gravitational Waves from Inflation}},  {\em JCAP} {\bf 09} (2012)
  017, [\href{http://arxiv.org/abs/1203.4663}{{\tt 1203.4663}}].

\bibitem{Assadullahi:2009nf}
H.~Assadullahi and D.~Wands, {\it {Gravitational waves from an early matter
  era}},  {\em Phys. Rev. D} {\bf 79} (2009) 083511,
  [\href{http://arxiv.org/abs/0901.0989}{{\tt 0901.0989}}].

\bibitem{Alabidi:2013lya}
L.~Alabidi, K.~Kohri, M.~Sasaki, and Y.~Sendouda, {\it {Observable induced
  gravitational waves from an early matter phase}},  {\em JCAP} {\bf 05} (2013)
  033, [\href{http://arxiv.org/abs/1303.4519}{{\tt 1303.4519}}].

\bibitem{Kovetz:2017rvv}
E.~D. Kovetz, {\it {Probing Primordial-Black-Hole Dark Matter with
  Gravitational Waves}},  {\em Phys. Rev. Lett.} {\bf 119} (2017), no.~13
  131301, [\href{http://arxiv.org/abs/1705.09182}{{\tt 1705.09182}}].

\bibitem{Kohri:2018awv}
K.~Kohri and T.~Terada, {\it {Semianalytic calculation of gravitational wave
  spectrum nonlinearly induced from primordial curvature perturbations}},  {\em
  Phys. Rev. D} {\bf 97} (2018), no.~12 123532,
  [\href{http://arxiv.org/abs/1804.08577}{{\tt 1804.08577}}].

\bibitem{Inomata:2019ivs}
K.~Inomata, K.~Kohri, T.~Nakama, and T.~Terada, {\it {Enhancement of
  Gravitational Waves Induced by Scalar Perturbations due to a Sudden
  Transition from an Early Matter Era to the Radiation Era}},  {\em Phys. Rev.
  D} {\bf 100} (2019) 043532, [\href{http://arxiv.org/abs/1904.12879}{{\tt
  1904.12879}}]. [Erratum: Phys.Rev.D 108, 049901 (2023)].

\bibitem{Domenech:2020ssp}
G.~Dom\`enech, C.~Lin, and M.~Sasaki, {\it {Gravitational wave constraints on
  the primordial black hole dominated early universe}},  {\em JCAP} {\bf 04}
  (2021) 062, [\href{http://arxiv.org/abs/2012.08151}{{\tt 2012.08151}}].
  [Erratum: JCAP 11, E01 (2021)].

\bibitem{Kibble:1982dd}
T.~W.~B. Kibble, G.~Lazarides, and Q.~Shafi, {\it {Walls Bounded by Strings}},
  {\em Phys. Rev. D} {\bf 26} (1982) 435.

\bibitem{Preskill:1992ck}
J.~Preskill and A.~Vilenkin, {\it {Decay of metastable topological defects}},
  {\em Phys. Rev. D} {\bf 47} (1993) 2324--2342,
  [\href{http://arxiv.org/abs/hep-ph/9209210}{{\tt hep-ph/9209210}}].

\bibitem{Dunsky:2021tih}
D.~I. Dunsky, A.~Ghoshal, H.~Murayama, Y.~Sakakihara, and G.~White, {\it {GUTs,
  hybrid topological defects, and gravitational waves}},  {\em Phys. Rev. D}
  {\bf 106} (2022), no.~7 075030, [\href{http://arxiv.org/abs/2111.08750}{{\tt
  2111.08750}}].

\bibitem{Babichev:2021uvl}
E.~Babichev, D.~Gorbunov, S.~Ramazanov, and A.~Vikman, {\it {Gravitational
  shine of dark domain walls}},  {\em JCAP} {\bf 04} (2022), no.~04 028,
  [\href{http://arxiv.org/abs/2112.12608}{{\tt 2112.12608}}].

\bibitem{Babichev:2023pbf}
E.~Babichev, D.~Gorbunov, S.~Ramazanov, R.~Samanta, and A.~Vikman, {\it
  {NANOGrav spectral index \ensuremath{\gamma}=3 from melting domain walls}},
  {\em Phys. Rev. D} {\bf 108} (2023), no.~12 123529,
  [\href{http://arxiv.org/abs/2307.04582}{{\tt 2307.04582}}].

\bibitem{Schroder:2024gsi}
T.~Schr\"oder and R.~Brandenberger, {\it {Embedded domain walls and electroweak
  baryogenesis}},  {\em Phys. Rev. D} {\bf 110} (2024), no.~4 043516,
  [\href{http://arxiv.org/abs/2404.13035}{{\tt 2404.13035}}].

\bibitem{Lu:2023mcz}
B.-Q. Lu, C.-W. Chiang, and T.~Li, {\it {Clockwork axion footprint on nanohertz
  stochastic gravitational wave background}},  {\em Phys. Rev. D} {\bf 109}
  (2024), no.~10 L101304, [\href{http://arxiv.org/abs/2307.00746}{{\tt
  2307.00746}}].

\bibitem{Gouttenoire:2023ftk}
Y.~Gouttenoire and E.~Vitagliano, {\it {Domain wall interpretation of the PTA
  signal confronting black hole overproduction}},  {\em Phys. Rev. D} {\bf 110}
  (2024), no.~6 L061306, [\href{http://arxiv.org/abs/2306.17841}{{\tt
  2306.17841}}].

\bibitem{Blasi:2023sej}
S.~Blasi, A.~Mariotti, A.~Rase, and A.~Sevrin, {\it {Axionic domain walls at
  Pulsar Timing Arrays: QCD bias and particle friction}},  {\em JHEP} {\bf 11}
  (2023) 169, [\href{http://arxiv.org/abs/2306.17830}{{\tt 2306.17830}}].

\bibitem{Antoniadis:2022pcn}
J.~Antoniadis {\em et~al.}, {\it {The International Pulsar Timing Array second
  data release: Search for an isotropic gravitational wave background}},  {\em
  Mon. Not. Roy. Astron. Soc.} {\bf 510} (2022), no.~4 4873--4887,
  [\href{http://arxiv.org/abs/2201.03980}{{\tt 2201.03980}}].

\bibitem{Chiang:2020aui}
C.-W. Chiang and B.-Q. Lu, {\it {Testing clockwork axion with gravitational
  waves}},  {\em JCAP} {\bf 05} (2021) 049,
  [\href{http://arxiv.org/abs/2012.14071}{{\tt 2012.14071}}].

\bibitem{Lu:2024ngi}
B.-Q. Lu, C.-W. Chiang, and T.~Li, {\it {Primordial black hole from domain wall
  fluctuations}},  \href{http://arxiv.org/abs/2409.09986}{{\tt 2409.09986}}.

\bibitem{Lu:2024szr}
B.-Q. Lu, C.-W. Chiang, and T.~Li, {\it {A Common Origin for Nano-Hz
  Gravitational Wave Background and Black Hole Merger Events}},
  \href{http://arxiv.org/abs/2409.10251}{{\tt 2409.10251}}.

\bibitem{Saito:2008jc}
R.~Saito and J.~Yokoyama, {\it {Gravitational wave background as a probe of the
  primordial black hole abundance}},  {\em Phys. Rev. Lett.} {\bf 102} (2009)
  161101, [\href{http://arxiv.org/abs/0812.4339}{{\tt 0812.4339}}]. [Erratum:
  Phys.Rev.Lett. 107, 069901 (2011)].

\bibitem{Saito:2009jt}
R.~Saito and J.~Yokoyama, {\it {Gravitational-Wave Constraints on the Abundance
  of Primordial Black Holes}},  {\em Prog. Theor. Phys.} {\bf 123} (2010)
  867--886, [\href{http://arxiv.org/abs/0912.5317}{{\tt 0912.5317}}]. [Erratum:
  Prog.Theor.Phys. 126, 351--352 (2011)].

\bibitem{DeLuca:2019ufz}
V.~De~Luca, G.~Franciolini, A.~Kehagias, and A.~Riotto, {\it {On the Gauge
  Invariance of Cosmological Gravitational Waves}},  {\em JCAP} {\bf 03} (2020)
  014, [\href{http://arxiv.org/abs/1911.09689}{{\tt 1911.09689}}].

\bibitem{Inomata:2019yww}
K.~Inomata and T.~Terada, {\it {Gauge Independence of Induced Gravitational
  Waves}},  {\em Phys. Rev. D} {\bf 101} (2020), no.~2 023523,
  [\href{http://arxiv.org/abs/1912.00785}{{\tt 1912.00785}}].

\bibitem{Yuan:2019fwv}
C.~Yuan, Z.-C. Chen, and Q.-G. Huang, {\it {Scalar induced gravitational waves
  in different gauges}},  {\em Phys. Rev. D} {\bf 101} (2020), no.~6 063018,
  [\href{http://arxiv.org/abs/1912.00885}{{\tt 1912.00885}}].

\bibitem{Vilenkin:1984ib}
A.~Vilenkin, {\it {Cosmic Strings and Domain Walls}},  {\em Phys. Rept.} {\bf
  121} (1985) 263--315.

\bibitem{Nambu:1990ez}
Y.~Nambu, H.~Ishihara, N.~Gouda, and N.~Sugiyama, {\it {Anisotropies of the
  cosmic background radiation by domain wall networks}}, .

\bibitem{Kaiser:1990xe}
N.~Kaiser and J.~A. Peacock, {\it {Power spectrum analysis of one-dimensional
  redshift surveys}},  {\em Astrophys. J.} {\bf 379} (1991) 482--506.

\bibitem{Hiramatsu:2013qaa}
T.~Hiramatsu, M.~Kawasaki, and K.~Saikawa, {\it {On the estimation of
  gravitational wave spectrum from cosmic domain walls}},  {\em JCAP} {\bf 02}
  (2014) 031, [\href{http://arxiv.org/abs/1309.5001}{{\tt 1309.5001}}].

\bibitem{Martins:2016book}
C.~Martins, {\em {Defect Evolution in Cosmology and Condensed Matter}}.
\newblock Springer, 2016.

\bibitem{Martins:2016ois}
C.~J. A.~P. Martins, I.~Y. Rybak, A.~Avgoustidis, and E.~P.~S. Shellard, {\it
  {Extending the velocity-dependent one-scale model for domain walls}},  {\em
  Phys. Rev. D} {\bf 93} (2016), no.~4 043534,
  [\href{http://arxiv.org/abs/1602.01322}{{\tt 1602.01322}}].

\bibitem{Baumann:2007zm}
D.~Baumann, P.~J. Steinhardt, K.~Takahashi, and K.~Ichiki, {\it {Gravitational
  Wave Spectrum Induced by Primordial Scalar Perturbations}},  {\em Phys. Rev.
  D} {\bf 76} (2007) 084019, [\href{http://arxiv.org/abs/hep-th/0703290}{{\tt
  hep-th/0703290}}].

\bibitem{Papanikolaou:2020qtd}
T.~Papanikolaou, V.~Vennin, and D.~Langlois, {\it {Gravitational waves from a
  universe filled with primordial black holes}},  {\em JCAP} {\bf 03} (2021)
  053, [\href{http://arxiv.org/abs/2010.11573}{{\tt 2010.11573}}].

\bibitem{Cai:2019cdl}
R.-G. Cai, S.~Pi, and M.~Sasaki, {\it {Universal infrared scaling of
  gravitational wave background spectra}},  {\em Phys. Rev. D} {\bf 102}
  (2020), no.~8 083528, [\href{http://arxiv.org/abs/1909.13728}{{\tt
  1909.13728}}].

\bibitem{Hiramatsu:2012sc}
T.~Hiramatsu, M.~Kawasaki, K.~Saikawa, and T.~Sekiguchi, {\it {Axion cosmology
  with long-lived domain walls}},  {\em JCAP} {\bf 01} (2013) 001,
  [\href{http://arxiv.org/abs/1207.3166}{{\tt 1207.3166}}].

\end{thebibliography}\endgroup
\end{document}